\documentclass[reprint,twocolumn,superscriptaddress,prx,longbibliography,groupedaddress,nofootinbib]{revtex4-2}

\usepackage{amsmath,amssymb}
\usepackage{graphicx}
\usepackage{dcolumn}
\usepackage{bm}
\usepackage{xr}
\usepackage[colorlinks = true,linktocpage=true]{hyperref}
\usepackage{tcolorbox}

\usepackage{graphicx} 
\usepackage{physics}  
\usepackage{xcolor}

\usepackage{amssymb}
\usepackage{amsmath}

\usepackage{hyperref}        
\hypersetup{colorlinks=true,breaklinks,
             linkcolor=black,urlcolor=black,
             anchorcolor=black,citecolor=black}

\newcommand*{\tran}{{\mkern-1.5mu\mathsf{T}}}		

\begin{document}

\markboth{Alvarado, et al.}{Optimal Control in Soft and Active Matter}
\title{Optimal Control in Soft and Active Matter}

\author{Jos\'e Alvarado}
\email{Corresponding author: alv@chaos.utexas.edu}
\affiliation{Center for Nonlinear Dynamics, Department of Physics, University of Texas, Austin, Texas, USA, 78712}
\author{Erin G.\ Teich}
\email{et106@wellesley.edu}
\affiliation{Department of Physics and Astronomy, Wellesley College, Wellesley, Massachusetts, USA, 02481}
\author{David A.\ Sivak}
\email{dsivak@sfu.ca}
\affiliation{Department of Physics, Simon Fraser University, Burnaby, British Columbia, Canada, V5A1S6}
\author{John Bechhoefer}
\email{Corresponding author: johnb@sfu.ca}
\affiliation{Department of Physics, Simon Fraser University, Burnaby, British Columbia, Canada, V5A1S6}




\begin{abstract}
Soft and active condensed matter represent a class of fascinating materials that we encounter in our everyday lives---and  constitute life itself. Control signals interact with the dynamics of these systems, and this influence is formalized in control theory and optimal control. Recent advances have employed various control-theoretical methods to design desired dynamics, properties, and functionality. Here we provide an introduction to optimal control aimed at physicists working with soft and active matter. We describe two main categories of control, feedforward control and feedback control, and their corresponding optimal control methods. We emphasize their parallels to Lagrangian and Hamiltonian mechanics, and provide a worked example problem. Finally, we review recent studies of control in soft, active, and related systems. Applying control theory to soft, active, and living systems will lead to an improved understanding of the signal processing, information flows, and actuation that underlie the physics of life.
\end{abstract}

\maketitle

\tableofcontents





\section{INTRODUCTION}

Living systems, from single cells to humans, perform a wide array of functional behaviors.
On organismal scales, tasks such as walking, jumping, foraging, hunting, and even scoring a goal in a soccer match require the orchestration of a plethora of physical effects for proper execution.
Similar orchestration occurs on cellular scales, in a similarly wide range of tasks including wound healing, division, shape change, and even goal-oriented locomotion.
The laws of physics suffice to describe the temporal evolution of a dynamical system and the macroscopic properties of an ensemble of microscopic constituents.
But is knowledge of the laws of physics sufficient to understand how living systems orchestrate function?
Answering this question is difficult, given that living systems are complex and inherently out of equilibrium.
However, recent advances in the field of active matter provide insights on the fundamental physics needed for such an understanding.

\textit{Active matter} describes a class of inherently nonequilibrium systems, where molecular-scale mechanochemical activity self-organizes to yield emergent macroscopic phenomena.
Active matter is not only a useful platform to study fundamental nonequilibrium physics;
models from active-matter frameworks have successfully been extended to describe several kinds of biological phenomena, such as collective swarming, contractility, cell locomotion, and embryogenesis.
Active matter therefore offers researchers a class of tractable systems that incorporate some of the core properties of living systems.

Naturally, active-matter descriptions alone are not sufficient to completely describe the orchestration of all biological function: cells and organisms routinely make decisions and execute these decisions by influencing their state with \textit{control signals}.
Perhaps the most well-suited formalism for understanding control signals lies in \textit{control theory}, which describes how control signals influence the dynamics of a system.

Control signals are ubiquitous in biology.
In the intracellular environment, enzymatic activity is regulated by post-translational modifications (e.g., phosphorylation); meanwhile, chemical signals from the environment impart information on cell decision-making processes.
On organismal scales, muscle and neuronal excitatory and inhibitory activity is controlled by action potentials.
In all cases, these control signals can directly affect microscopic mechanochemical activity and thus self-organization and macroscopic behavior.
Therefore, studying how control signals interact with the laws of physics is an essential component to understanding orchestration of active self-assembly and ultimately biological function.

Which control signal is the ``right'' one to steer an active system to exhibit a desired function?
This problem is not immediately straightforward.
One branch of control theory, \textit{optimal control}, provides a library of methods that aim to solve this problem.
Existing review articles have covered applications of optimal control to soft and active systems \cite{Liu2023, Solomon2018, Takatori-2025-Annu.Rev.Condens.MatterPhys., Paulson2015, Tang2022, N.McDonald-2023-SoftMatter}.
However, control theory is often not a part of physics curricula, and physicists may lack the necessary background to get started.
The focus of this review article is to introduce optimal control to an audience of physicists working with soft and active matter who are unfamiliar with these frameworks.
We provide an introduction to \textit{feedforward optimal control}, which rests on a detailed physical model of the system to be controlled.
We contrast these strategies with \textit{feedback optimal control}, which instead relies on information obtained from system observations.
Combining feedforward and feedback allows for better tradeoffs between performance and robustness.
Furthermore, we briefly discuss the relation between optimal control and reinforcement learning.
Finally, we provide an overview of recent studies of optimal control in soft, active, and related systems.
But first, we provide a brief overview of the fundamentals of control theory. 

\section{CONTROL}

\subsection{Control Theory}

Although control theory is a well-developed discipline with roots in engineering and applied mathematics, and although it has increasing applications within physics, the subject is rarely taught in standard physics curricula. Here, we present a few basic ideas, leading to a brief description of optimal control and its place within the broader field. For an in-depth treatment, see \cite{bechhoefer2021control}.

\begin{figure} [ht]
    \begin{center}
	   \includegraphics[width=\columnwidth]
       {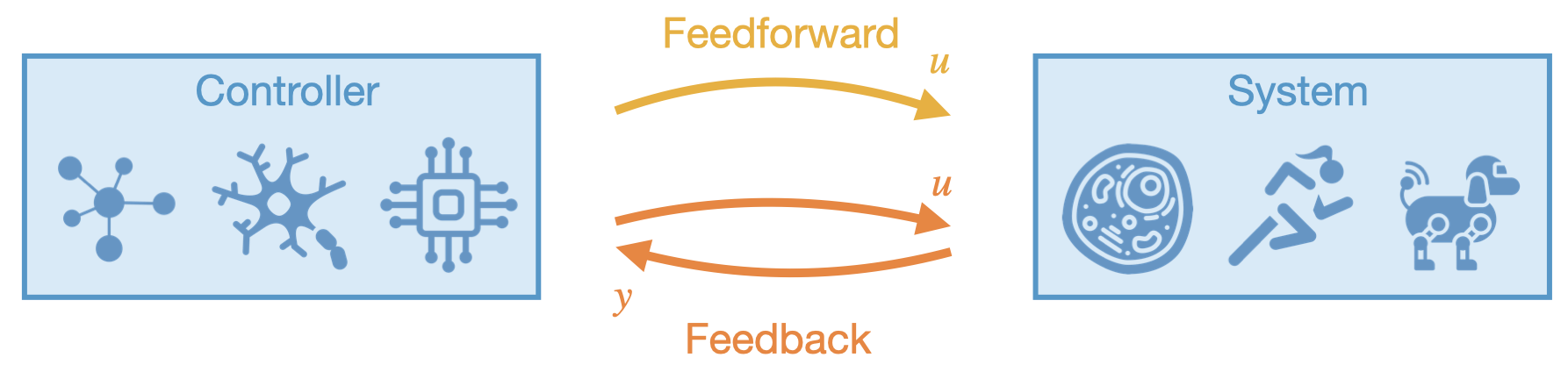}
	   \caption{Schematic of a controller performing feedforward and feedback control on a dynamical system.}  	
	   \label{fig:controller-system}
    \end{center}
\end{figure}

We begin with the notions of a physical system (industrial ``plant" in the engineering literature) and a controller.  For us, they will informally be dynamical systems that interact with each other.  In the simpler situation of \textit{open-loop} control, the controller influences the physical system, but there is negligible back-action on the physical system; in the more complicated \textit{closed-loop} control, the influence goes both ways (Fig.~\ref{fig:controller-system}).
The open-loop controller operates by \textit{feedforward}, in that the control is independent of the dynamical state (``state,'' for short) of the physical system.  The closed-loop controller operates by \textit{feedback}, in which the control action is a function of the system state (now and, perhaps, in the past).

We focus on continuous-time dynamical systems governed by differential equations.  In control theory, it is important to explicitly specify both the inputs to a dynamical system and its outputs.  The former describes how the system is controlled externally, and the latter describes the measurements available to monitor the system and allow feedback.  More formally, a system is characterized by a time-dependent $n_x$-dimensional state vector $x(t)$ that evolves according to nonlinear dynamics $\dot{x}~=~f(x,u)$, where $u(t)$ is an $n_u$-dimensional vector of input signals and $f(\cdot,\cdot)$ is a nonlinear function of dimension $n_x$.  The system outputs are characterized by the $n_y$-dimensional vector $y(t)$, which is related to the system state via a nonlinear function $y = h(x)$.  Abstractly, a system transforms an input signal $u(t)$ into an output signal $y(t)$. 

While the above description is deterministic, most systems are affected by noise, which can enter in several ways.  Soft- and active-matter systems are usually sensitive to thermal noise and may also be affected by the fluctuations in a nonequilibrium environment.  When under control, both outputs and inputs also contribute noise.  The former 
is known as \textit{measurement noise} and is often modeled by an additive Gaussian noise term.  


\subsection{Goals of Control}
\label{sec:goals}

The general goal of control is to ``improve" the dynamics in some way.
In this review, we focus on three simple cases.
The first is \textit{state-to-state transformation}, where the goal is to go from $x(0)~=~x_0$ to $x(\tau)~=~x_\tau$ without placing demands on how to get from start to end.
In this case, one can consider the state $x$ to represent a single particle; more broadly, $x$ could refer to an entire system of particles, allowing one to view state-to-state transformation as switching phases, material properties, or dynamical behavior.
A second case is \textit{tracking}, where the system state vector $x(t)$ should follow a desired time-dependent trajectory $x^*(t)$ as closely as possible.
That is, we try to keep $x(t)$ in a ``tube'' centered on a curve in state space.
To realize these goals, a controller needs a good model of the system.
As detailed in §\ref{sec:optimal}, feedforward optimal-control methods determine the optimal control signal $u^*(t)$ that achieves these goals; however, in practice, not all models agree perfectly with experiment.
Information from sensors in the system can be sent to the controller in a feedback loop, enabling a third control goal, \textit{regulation}, where the internal state should be kept constant, e.g., $x(t) = x^*$.
This was historically perhaps the first goal of control.
The idea is to fix a system's state in the face of perturbations; e.g., we can regulate temperature (of a physical sample, of our body, etc.) against environmental variations.
Apart from these three goals, additional ones include stabilizing an unstable equilibrium, suppressing chaotic motion, or creating collective states such as synchronized oscillators.

\subsection{Obstacles}
\label{sec:obstacles}

There are typical obstacles that make these goals of control hard to achieve.  We already mentioned that the environment and measurements may be noisy, and the resulting stochasticity must be filtered or otherwise compensated for.  There are also limits associated with the system itself.  Some of these depend on fundamental thermodynamics and statistical physics and trace back to the requirement in control for the acquisition and processing of information~\cite{parrondo2015thermodynamics}.  But more commonly, control is in practice limited by other issues.  For example, the control variables (or \textit{actuator}, in the engineering jargon) have physical limits: a motor has a speed limit, light intensity can only be so high in an experiment, and so on.  More subtly, a control actuator may be coupled to a system in a way that limits its \textit{controllability}, the set of states that can be attained. The notion implicitly assumes that a control signal can be arbitrarily large.  Accounting for a finite range of control inputs leads to the notion of \textit{reachability}. These notions of controllability and reachability have analogs for inputs from sensors, suggesting a notion of \textit{observability}.  Note that it may be possible to only control or observe a subspace of the entire state space.


Other constraints and costs influence the ability to control.  For example, there may be physical constraints: one may want to control motion to avoid a particular region of space, or temperature should be controlled to not exceed some limit.  Finally, even when a controller can reach a particular state, there are typically costs associated with the use of control itself; for example, the energy required for the control may be significant and need to be minimized.

\section{FEEDFORWARD OPTIMAL CONTROL}
\label{sec:optimal}

In feedforward control, one pre-plans control based on the expected behavior of the system (incorporating prior knowledge), and the control is enacted without ongoing information (feedback) about the current system behavior. 

\subsection{Feedforward Control}
\label{sec:feedforward}

If you know (reasonably) well the dynamics of the system you wish to control and if you know what you would like it to do, feedforward control is the most efficient approach. 

\textit{Inverse control}.   A particularly simple case occurs when the dynamics are invertible.  That is, we can ``solve" or rewrite the dynamics $\dot{x}~=~f(x,u)$ in the form $u(t) = F(x)$ for some function $F(\cdot)$.  For example, the system $\dot{x} = -x + u$ can be solved as $u(t) = \dot{x}(t) + x(t)$.  To make $x(t)$ track some desired trajectory $x^*(t)$, we simply invert the dynamical relation and set $u^*(t) = \dot{x}^*(t) + x^*(t)$.  Notice that we do not have to solve the system's differential equation; we just need to take derivatives of $x^*(t)$.

\begin{figure} [ht]
    \begin{center}
	   \includegraphics
       [width=\columnwidth]
       {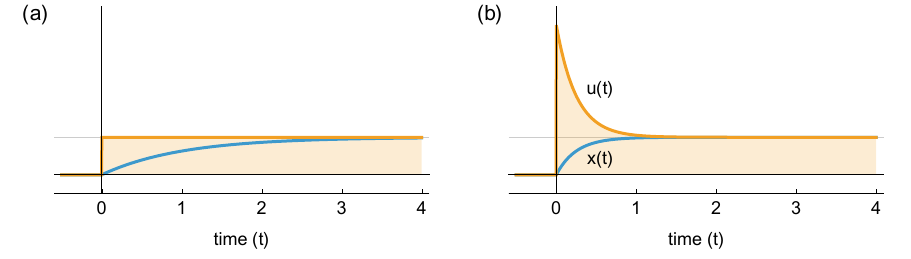}
	   \caption{Feedforward control of a first-order linear system.  (a) Response $x(t)$ to a step input $u(t) = \theta(t)$.  (b) The higher-amplitude input $u^*(t)$ produces a faster response $x^*(t)$.}  	
	   \label{fig:ff-example}
    \end{center}
\end{figure}

As a first-order, one-dimensional, linear example, the normal response to a step-function input $u(t) = \theta(t)$ is $x(t) = 1-{\rm e}^{-t}$ (Fig.~\ref{fig:ff-example}a).  To ``speed up" the response to $x^*(t) = 1-{\rm e}^{-\lambda t}$, with $\lambda > 1$, we substitute $x^*(t)$ into the inverted dynamical relation to find $u^*(t) = (\lambda-1){\rm e}^{-\lambda t}+1$ (Fig.~\ref{fig:ff-example}b).  The maximum required value of $u(t)$ has increased by a factor $\lambda$; this extra power (in a physical realization) is the price to pay for accelerated motion.  Such speed-energy tradeoffs are characteristic of the thermodynamics of control processes.

There are obvious limitations on the ability to carry out such inverse control.
First, the dimension of the input must equal or exceed that of the state; that is, we need one input for every degree of freedom.  In practice, the number of degrees of freedom greatly exceeds the number of control parameters.  E.g., a piston controls the volume of a chamber; the single control parameter controls the overall pressure, but obviously cannot control the detailed motion of the gas molecules in the chamber.
Second, the maximum value of an actuator, the physical device affecting or controlling the system, is always limited to some finite value.  In the above example, we must choose $\lambda = $ max[$u(t)] \le |u|_{\rm max}$.

This inverse method allows one to find a control without integrating the equations of motion, but the number $n_u$ of control parameters must equal the number of degrees of freedom.
More generally, for systems with the geometric property of \textit{differential flatness}, one can find analogous relations for a subset of $n_u$ coordinates~\cite{fliess1995flatness}. For example, a quadrotor drone is a rigid object with six degrees of freedom but only four control parameters, the rotation rates of its four rotors; nonetheless, the four controls can be derived from a desired trajectory ($x^*,y^*,z^*$) and heading (yaw angle)~\cite{mellinger2011minimum}.

\subsection{Optimal Control}

Optimal control provides a systematic way to design controllers and is increasingly popular in more complex applications.  The basic concept is to define a scalar \textit{cost function} $J$ that penalizes poor control performance and excessive control use.\footnote{
Economists and computer scientists are more optimistic and instead define a \textit{reward function}.  Since $-J$ can serve as a reward, the two are equivalent.}
Lower costs are better, and $J$ is bounded from below, conventionally by zero, which would reflect ``perfect" control.\footnote{
Never confuse ``optimal" with ``good":  Optimal control finds the control that minimizes a cost function,
but a poorly chosen cost function can lead to bad, even disastrous results.  As Aesop observed long ago, ``We would often be sorry if our wishes were gratified"~\cite{jacobs1894the}.}

A typical cost function for a protocol that starts at $t=0$ and ends at $t=\tau$ is of the form $J = \varphi(x_\tau,\tau) + \int_0^\tau \dd{t} L(x(t),u(t))$. Here $\varphi(x_\tau,\tau)$ penalizes deviations from a desired end state and (if there is a term proportional to $\tau$) a longer time $\tau$. The \textit{running cost} $L(x,u)$ quantifies deviations of the trajectory from a desired $x^*(t)$ and is an increasing function of $u$, to reduce control usage.  A typical form is $L = \tfrac{1}{2}(x^\tran Q x + u^\tran R u)$.
The matrices $Q$ and $R$ are positive semidefinite and usually diagonal, with each element penalizing deviations for a particular component of the system state or control vector, respectively. 

If the system has fewer degrees of control than degrees of freedom (\textit{underactuated}), one can still drive it from one state to another, but it is no longer possible to invert the dynamics to find a control for an arbitrary trajectory.  In such a case, the $Q$ term is absent and one evaluates the protocol solely on the control usage (via the $R$ term) and the accuracy of the end state (via $\varphi$ or possibly by imposing a strict boundary condition).  For example, a cart on a track carrying a pendulum has two degrees of freedom but only one control parameter (the force on the cart); nonetheless, one can find a control force $u(t)$ that takes the system from ``Down" ($x=\theta=0$) to ``Up" ($x=0$, $\theta=\pi$).

\subsection{Optimal Feedforward Control From a Given Initial State}

The most straightforward approach to optimal control is to seek a feedforward control $u(t)$ that drives the system from a given initial state $x(0) = x_0$ towards a desired final state $x(\tau) = x_\tau$.  Here, we simplify by fixing the protocol end time $\tau$ and by demanding that $x(\tau) = x_\tau$ (rather than a soft final constraint), implying $\varphi=0$.  This leaves the integrated running cost $L(x,u)$; we seek to minimize $J$, subject to the constraint that a known dynamics $\dot{x}=f(x,u)$ is obeyed.  This constrained optimization problem can be formulated as an unconstrained optimization of a modified cost,
\begin{align}
	J' = \int_0^\tau \dd{t} \left( L(x,u) + \lambda^\tran(t)[f(x,u)-\dot{x}] \right) \,.
\end{align}  
The Lagrange multiplier $\lambda(t)$ has the same dimension as $x(t)$ and becomes a dynamical variable in its own right, known as the \textit{adjoint} or \textit{co-state}.

One can minimize $J'$ with respect to the unknown control $u(t)$ using the calculus of variations.  This gives 
$2n_x + n_u$ total
Euler-Lagrange equations for $x(t)$, $\lambda(t)$, and $u(t)$,
with structure closely analogous to 
Lagrangian mechanics; in particular, the Lagrange multiplier $\lambda(t)$ plays the role of a conjugate momentum with respect to the state $x(t)$.  This motivates a Hamiltonian formulation, with Hamiltonian $H(x,\lambda,u) = L + \lambda^\tran \dot{x} = L + \lambda^\tran f$.  (Here the sign of the Lagrange multiplier is opposite to that taken in classical mechanics. This is the usual convention in control-theory texts.)  
The Euler-Lagrange equations become Hamilton's equations, 
\begin{subequations}  \label{eq:ham}
\begin{alignat}{2}
	\dot{x} &= (\partial_\lambda H)^\tran &\quad &= f \,,  \label{eq:ham1} \\
		\dot{\lambda} &= -(\partial_x H)^\tran 
		&	&= -(\partial_x f)^\tran \lambda - (\partial_x L)^\tran \,, \label{eq:ham2} \\
		0 &= (\partial_u H)^\tran  &	&= (\partial_uf)^\tran \lambda +(\partial_u L)^\tran \,. \label{eq:ham3}
\end{alignat}
\end{subequations}
The transpose operations arise because $\partial_x$ generates a row vector in our notation.  The first equation, for the state $x$, reproduces the equations of motion, $\dot{x}=f(x,u)$.  It obeys a condition at $t=0$.  The second equation, for the adjoint $\lambda$, is new and obeys at $t=\tau$ a condition on either the adjoint or the state.  Intuitively, $\lambda(t)$ captures the notion of planning:  what needs to be done now in order to achieve a desired state in the future. The third equation, \eqref{eq:ham3}, gives algebraic relations (one per control component) that act as compatibility conditions relating state, adjoint, and control variables.  Together, Eqs.~\eqref{eq:ham} define a two-point boundary-value ordinary-differential-equation problem in time, with $n_x$ conditions at $t=0$ and $n_x$ at $t=\tau$.  This can be solved by shooting, in which the $x(t)$ equation is integrated forward in time and the $\lambda(t)$ equation is integrated backwards in time, and then one iterates to convergence.  Finding a sufficiently close initial guess can be tricky, though, motivating consideration of more sophisticated pseudospectral collocation methods with better convergence properties~\cite{ross2012review}. 

\subsection{Pontryagin Minimum Principle (PMP)}
\label{sec:pmp}

An extension of these equations was derived in the late 1950s by Pontryagin.  The equation for $u(t)$ corresponds to finding a minimum (or, at least, extremum) of $J'$.  But physically realizable controls have bounds on $u(t)$, and the minimum may lie outside those bounds.   In these cases, $H$ is minimized with $u(t)$ ``pinned" to the boundary at time $t$.  

Pontryagin showed that these issues can be dealt with if one first defines a \textit{control Hamiltonian} $\mathcal{H}(x,\lambda) = \inf_{u \, \in \, \mathcal{U}} H(x,\lambda,u)$.  
The minimization corresponds to Eq.~\eqref{eq:ham3} for solutions in the interior of the allowed controls $\mathcal{U}$ and otherwise lies on the boundary.
His theorem, the Pontryagin Minimum Principle (PMP), provides necessary conditions that an optimal solution must obey.\footnote{If $H$ is defined as $\lambda^\tran f - L$, this becomes the Pontryagin Maximum Principle.}
Once the minimization is carried out, the control Hamiltonian obeys exact analogs of Hamilton's equations:  $\dot{x} = (\partial_\lambda \mathcal{H})^\tran$ and $\dot{\lambda} = -(\partial_x \mathcal{H})^\tran.$   
 
The running cost $L$ can be independent of $u$, for example if we care only about the accuracy of a state or seek to transform a state as quickly as possible (minimum-time control).  If $u$ also linearly affects the equations of motion (e.g., under a driving force), then $H$ is linear in $u$ and there is no minimum at finite $u$: the control $u(t)$ always lies on the boundary.  If one-dimensional, control alternates between lower and upper bounds (\textit{bang-bang control}).  For example, to drive from one point to another in minimum time, one should accelerate as fast as possible until some intermediate time and then brake as hard as possible the rest of the way.  There are also biological examples where bang-bang control has been found, including development of intestinal crypts~\cite{itzkovitz2012optimality} and control of protein aggregation~\cite{michaels2019optimal,dear2021feedback}

As an example of optimization that is not constrained by boundaries, consider the swing-up of a pendulum via an applied torque, with $\ddot{\theta} + \sin \theta = u$.  
(The second-order form in terms of $\theta(t)$ 
is simpler than the equivalent two first-order equations)  
The equation for the control gives $u= -\lambda$ and that for the adjoint, after eliminating $u$, is $\ddot{\lambda} + \lambda \cos \theta= 0$.  The initial condition 
in the Down state 
is $\theta = \dot{\theta} = 0$ at $t=0$.
The final condition 
in the Up state
is $\theta = \pi$ and $\dot{\theta}=0$.
We thus have four first-order equations and four boundary conditions, giving a well-defined mathematical problem.  Such problems can almost never be solved analytically.  

The numerical solution of Hamilton's equations is delicate: although the Hamiltonians of classical mechanics describe systems with no dissipation and are invariant under time reversal, the dynamical equations for control problems can be dissipative---and usually are.  The structure of Hamilton's equations then implies that every dynamical eigenvalue with a negative real part (corresponding to stable motion when the equation is propagated forward in time) has a positive partner with equal but positive real part ($\lambda(t)$ dynamics that are stable when integrated backwards in time).  But we actually integrate both equations, meaning that nonzero eigenvalues lead to unstable growth in some variables, whichever way the equations are integrated.  Since protocol times are finite, there is no divergence; nonetheless, the differential equations are often stiff and thus hard to solve.

\section{FEEDBACK OPTIMAL CONTROL}

Feedforward control is pre-planned and executed without real-time information about the system behavior; \textit{feedback control}, by contrast, uses real-time measurements to modify the originally planned control in response to current system behavior.

\subsection{Feedback Control}
\label{sec:feedback}

Feedback control can compensate for uncertainties resulting from external disturbance or from poorly known system dynamics.  Adding sensors to measure and infer the system state allows control of unexpected deviations.  Constructing feedback generally requires two steps: (i) from sensor measurements, estimate the current system state; (ii) compute a control action to ``correct" for deviations between the desired and inferred system state.

The simplest feedback methods are heuristic and combine these two steps, going directly from the observations $y(t)$ to response $u(t)$.  The most popular algorithm, by far, is proportional-integral-derivative (PID) control, where the feedback response $u(t)$ is a function of the \textit{error} $e(t) \equiv r(t) - y(t)$ between the desired \textit{reference signal} $r(t)$ and the measurement $y(t)$.\footnote{The error's sign is opposite from what one might usually define.  To understand the reason, consider the simplest form of feedback, proportional control, where the control signal is proportional to the error, $u(t) = K_p e(t)$, for \textit{proportional gain} $K_p$.  The sign means that a negative error generates a positive response.}
For example, consider regulating a system's temperature.  Here, the reference $r(t)=r^*$ is constant, equaling the \textit{set point}.  If the system temperature is below the set point, a positive control signal implies turning on a heater to warm the system up.  This \textit{negative feedback} is typical when the goal is to stabilize a system.

More generally, the PID algorithm can be written
\begin{align}
	u(t) = K_p e(t) + K_i \int^t \dd{t'} e(t') + K_d \dv{e(t)}{t} \,.
\end{align} 
The first term is the proportional control (P) already mentioned.  The second term describes \textit{integral feedback} (I) with a strength $K_i$ necessary to compensate for a steady (constant) perturbation.  This plays a crucial role in many biological processes, for example in bacterial chemotaxis, where a bacterium moves along a concentration gradient with a response that is independent of the local absolution concentration~\cite{yi2000robust}.  The third term describes \textit{derivative feedback} (D) and responds strongly to sudden changes in the error.  Intuitively, if the system starts moving away from a set point, one can infer that there has been a large perturbation and can, in anticipation, start to compensate for it before the full effects have been felt.  Of course, the inference could be mistaken---noise can also make a signal seem to change suddenly---and it is important to have good filtering algorithms that distinguish between noise and signal in a measured quantity and can infer unmeasured states.  (For example, often some quantity is measured, but its derivative, which is also a required state variable, must be inferred.)

The combination of P and I and D is very effective, and simple applications of feedback mostly use this technique.  The values of the three feedback gains $K_p$, $K_i$, and $K_d$ are chosen as functions of system dynamics.  For a system with nonlinear dynamics, local linearization around a fixed set point will determine appropriate choices.  When changing the set point or tracking a moving reference, the simplest generalization is to adopt local values of feedback gains corresponding to local linearization of the dynamics.  This technique of \textit{gain scheduling} works as long as the nonlinearities are weak or the velocity through state space is sufficiently small.  For more intrinsically nonlinear control problems, more sophisticated techniques are needed.

For systems that are linear (or nearly so), there are many other heuristic techniques.  One class, suitable for time-invariant systems, uses Laplace (or Fourier) transforms to design the controller by tailoring the complex frequency response of the controller as a function of the system response (\textit{transfer function}).  Other techniques work in the time domain and rely on linear-algebra techniques to manipulate the eigenvalues describing the system's response and closed-loop dynamics.  Such techniques can be more easily extended to time-dependent and nonlinear systems but are less intuitive than frequency-domain methods.

The methods above are effective for simple systems with sufficiently numerous control parameters and sensors and sufficiently small nonlinearities.  But the methods are somewhat heuristic and in practice lead to significant time needed for ``trial and error."  One particular problem is that the number of feedback gains to fix grows with the complexity of the overall system.  Loosely, PID control is effective for a system that has one dominant mode.

Systems with many independent modes need controllers with two or three parameters per mode.  To control a multimode system requires tuning many parameters.  The tuning can be empirical or based on measurement of the system dynamics.  But each mode is also characterized by at least two parameters, whose values must also be estimated. Therefore, although pure feedback control may not be suitable for more complex systems, it can be combined with optimal control.

\subsection{Hamilton-Jacobi-Bellman Equation (HJB)}
\label{sec:ff-fb}

In §\ref{sec:pmp}, we discussed the Pontryagin Minimum Principle. Richard Bellman developed an alternative approach around the same time as Pontryagin. 
This approach builds on the \textit{principle of optimality}~\cite{bellman1957dynamic}: The optimal cost at any time $t$ along a path from 0 to $\tau$ equals the optimal cost from 0 to $t$ plus the optimal remaining cost from $t$ to $\tau$.  I.e., optimization is based on a sequence of optimal decisions, one at each time.  Then, the method of \textit{dynamic programming} works backwards from the goal (end state) and, optimizing at each time, determines an optimal path.

Such logic leads to the \textit{Bellman equation}~\cite{bertsekas2017dynamic}, which in discrete time reads
\begin{align}
    J^*(x_k) = 
	\underset{\{u_k\}}{\rm min} \left[ L(x_k, u_k) + J^*(x_{k+1}) \right] \,,
\label{eq:optimal-bellman-discrete}
\end{align}
for running cost $L(x_k,u_k)$ at time $t_k$.  
The \textit{cost-to-go} $J(x_k,u_k)$ gives the cost starting at time $t_k$, assuming a sequence of controls $\{u_k,u_{k+1}, \ldots, u_N\}$, where time $t_N$ denotes the protocol end.  The \textit{optimal cost-to-go} $J^*(x_k)$ is the minimum cost, achieved by making the optimal control choice $\{u^*_k,u^*_{k+1}, \ldots, u^*_N\}$ at each time step.  

Taylor expanding the $J^*(x_{k+1})$ term about $x_k$ and $t_k$ and then taking the continuous-time limit leads to the \textit{Hamilton-Jacobi-Bellman} (HJB) equation,
\begin{align}	
    -\partial_t J^*(x,t) =  \min_{u} \left[ L(x,u) + \left( \partial_{x} J^* \right) \, f(x,u) \right]
    = \mathcal{H}(x, \partial_xJ^*)\,,
\label{eq:optimal-HJB}
\end{align}
which is analogous to the Hamilton-Jacobi equation for the action in classical mechanics.  Again, the equation is integrated backwards in time from the end state $x_\tau$ to give the optimal feedback control $u^*(x(t),t)$. 

By contrast, the PMP (§\ref{sec:pmp}) gave a solution for one particular $x(t)$.  Solving the HJB equation amounts to solving for all optimal paths, one for each initial state $x$.  The solution is robust to perturbations of any size because one can always apply the optimal course of action, even if a perturbation abruptly alters the system state.

The HJB equation can accommodate stochastic perturbations by considering the \textit{expected} optimal cost-to-go $\langle J^* \rangle$, leading to an extra term in Eq.~\ref{eq:optimal-HJB} proportional to $\partial_{xx}J^*$.  See \cite{kappen2005path} for a derivation and path-integral solution.  Such thermal fluctuations are important in many soft- and active-matter applications. 

Despite its attractive features, the HJB equation is an $n_x$-dimensional system of nonlinear partial differential equations that is typically difficult to solve for even moderate state-space dimension.  The need to minimize over $u$ at each time makes even a numerical solution difficult in most cases.  The PMP is a $2n_x$-dimensional set of ordinary differential equations that is easier and much faster to solve numerically.

A special case of the HJB that \textit{can} be solved is regulation about a reference state for linear dynamics and quadratic costs.  For this \textit{linear quadratic regulator} (LQR), the equations of motion are linearized about a fixed point $f(x,u)=0$.  If the steady-state values of $x$ and $u$ are set to zero, the deviations obey $\dot{x} = Ax+Bu$, for dynamical matrix $A=\partial_x f$ and input coupling $B = \partial_u f$ both evaluated at the fixed point.  One can then explicitly solve (example in Supplement) for a feedback controller $u_{\rm fb} = -Kx$, where $K(t)$ is a matrix of feedback gains.  The gains go to zero at the end of the protocol, since control at time $t$ yields benefits only later (for protocols that last indefinitely, the optimal gains are constant).  The problem of tuning feedback gains, as in PID control, is replaced by the need to choose matrices $Q$ and $R$ in the running cost $L$ that weight state and control deviations.  These coefficients ideally have a more intuitive meaning; in practice, they are often determined by trial and error.

We have assumed that all $n_x$ states are fully observable, but often the number of observations $n_y < n_x$.  For observations $y = h(x) \approx Cx$, with $C$ an output-coupling matrix, one must infer states $x(t)$ from past observations of $y$.  For linear dynamics perturbed by Gaussian noise and with Gaussian measurement errors, the optimal estimator, a \textit{Kalman filter}~\cite{kalman1960new}, has an explicit solution whose structure parallels the LQR feedback controller~\cite{bechhoefer2021control}.

The LQR solution is readily generalized to a tracking solution by letting $x(t) \to \delta x(t) = x(t) - r(t)$, for reference trajectory $r(t)$.  Even more interesting, one can derive an LQR solution to linear feedback about a feedforward solution.  I.e., first use the PMP to find a particular feedforward control $u^*(t)$ that generates a nominal trajectory $x^*(t)$.  Since such a solution is fragile to perturbations and to modeling errors, then consider linear deviations about the trajectory, to derive an additional stabilizing feedback control that keeps the solution near the nominal optimal control.  The main difference in the calculation is that the linear matrices $A(t)$, $B(t)$, and $C(t)$ describing dynamics, input, and output coupling are all time dependent; nonetheless, the feedback-control solution closely parallels the simpler LQR case (example in Supplement). 

Despite the generic difficulty in solving the HJB equation even numerically, 
there is a trick that leads to solutions for a useful class of dynamics, so-called \textit{control affine} nonlinear systems with dynamics $\dot{x} = f(x) + g(x) u + \xi$, for $f$ an arbitrary (smooth) nonlinear function of the state and $\xi$ Gaussian noise, possibly colored.  If the cost is in also quadratic in $u$, then a clever change of variables (``Cole-Hopf transformation") linearizes the stochastic HJB equation~\cite{kappen2005path}.  
Despite the seemingly restrictive requirements, this class includes a range of physically interesting systems. 
The method was recently applied to
shepherding a ``flock" of interacting active particles across a complex landscape~\cite{sinha2023optimal},
and deserves to be more widely exploited.  

Another promising and ambitious approach, the \textit{Hamiltonian bridge}, aims at solving the HJB equation for spatially extended nonlinear pattern-forming systems~\cite{krishnan2024hamiltonian}.  The proposed technique uses \textit{smoothed particle hydrodynamics}~\cite{monaghan2012smoothed} to expand Eulerian fields over a particle-like basis, where the ``particles" obey Langevin-like equations. Since many soft and active-matter applications are to extended systems (§\ref{sec:activeLCgels}), this technique is particularly promising.

As an alternative when system complexity precludes direct solution of the HJB equation, the combination of a feedforward solution to the nonlinear control problem with a feedback stabilization about that solution can ensure robustness to moderate perturbations.  
Larger perturbations that take the system outside a ``tube" surrounding the nominal solution will invalidate the linear assumptions of the feedback.  A solution in that case is to recalculate the feedforward-feedback solution all over again~\cite{bechhoefer2021control},
similar to the technique of \textit{model predictive control} (MPC)~\cite{rawlings2024model}.  
In MPC, one first solves for the feedforward control $u_{\rm ff}(t)$ and then executes a short segment of it, lasting a time $\Delta t$.  The state at $x(t+\Delta t)$ is measured and the feedforward control recomputed, starting at the new state.  The $x(t)$ dependence of the initial condition of each recalculation gives an implicit feedback.  

The main issue with MPC is that the calculations must be done in a time $\Delta t \ll t_{\rm dyn}$, for the fastest relevant dynamical timescale $t_{\rm dyn}$.
The combination of feedforward-feedback-MPC relaxes the time constraint of simple MPC.  
The limiting factor 
for the recalculation
is the ability to predict the new state, which even for a chaotic system can be done on a timescale typically several times longer than $t_{\rm dyn}$.

\subsection{Reinforcement Learning}

Although we focus on applications of optimal control that use models of system dynamics, there has recently been an explosion of interest in adapting data-driven methods to control problems.  These methods are used to learn control laws, system dynamics, and environmental characteristics ranging from the statistics of fluctuations to the presence of obstacles that must be avoided. 

One such method, \textit{reinforcement learning}~\cite{sutton2018reinforcement}, has shown spectacular success in many applications, beating human champions at games such as chess and Go~\cite{silver2018general}.  Reinforcement learning (RL) is particularly suitable for controlling dynamical systems~\cite{recht2019tour,brunton2022data}.  For example, RL can race drones over obstacle courses, again beating both human champions~\cite{kaufmann2023champion} and classic optimal-control algorithms~\cite{song2023reaching}.

Here, RL is a method to learn control laws (``policies") that minimize costs (``maximize rewards") for controlled interactions with the environment.  Each interaction is assigned a reward that is evaluated over a long-enough time that ultimate success can be determined.  
If a model of system dynamics is available, then model-dependent RL learns the control faster than if not.
For unknown dynamics, model-free RL typically uses neural networks with multiple hidden layers (\textit{deepRL}) to approximate both the system dynamics and the control law~\cite{brunton2022data}.  

The resulting strategies balance exploration (learn more about the environment and dynamics) against exploitation (head to the goal). 
RL solves dynamic-programming and HJB-type equations of optimal control by direct adaptation~\cite{sutton1992reinforcement}. 
Against its many advantages is the need to train the algorithm.  In practice, most of the training can be via simulation, with only small amounts of physical experimentation needed~\cite{kaufmann2023champion}.

\subsection{Example: Navigation of Swimmers}
\label{sec:swimmers}

As discussed in §\ref{sec:goals}, a common control goal is state-to-state transformation.  One of the simplest examples is the literal transport of a particle from one position to another.  The transport of a Brownian particle from one position in a fluid to another using minimal work was posed and solved using Euler-Lagrange equations~\cite{schmiedl2007optimal}, finding a feedforward control that had discontinuities at the start and end of the protocol (example in Supplement).  

A similar problem on the macroscopic scale is for a swimmer that moves at constant velocity in a freely chosen direction to move in minimal time between two positions through a fluid with specified flow field.  This \textit{navigation problem} was solved by Zermelo in 1931, two decades before the advances in optimal control in the 1950s~\cite{zermelo1931uber}.

The Zermelo solution has inspired a series of studies on the navigation of active swimmers.  The Zermelo solution was applied to a variety of flow fields that an active swimmer might encounter, also considering other costs such as the heat dissipated into the fluid during the protocol~\cite{liebchen2019optimal}.  The calculations were later generalized to account for hydrodynamic interactions between the swimmer and walls or obstacles~\cite{daddi2021hydrodynamics}, influencing the path taken even in a liquid with no external flow fields.

Another set of investigations derived heuristic rules for a swimmer facing a complex, fluctuating flow field~\cite{piro2022optimal}, that are nearly optimal but can be implemented by a swimmer knowing the noise-free optimal path and having only local information about deviations from that path.  The strategy is a heuristic version of the optimal local feedback rule discussed at the end of §\ref{sec:ff-fb}.  

Although optimal navigation was originally formulated as a minimum-time problem, one can also seek to efficiently explore a neighborhood around a swimmer.  Thus, one can consider, for different initial orientations, the set of points reached after a given time, the \textit{isochrone curve}~\cite{piro2021optimal}.  Following optimal paths turns out to be a more efficient way to explore space than naive, straight-line paths; however, the curves tend to be chaotic at long times, implying a sensitive dependence on initial orientation that is absent for the straight-line strategy~\cite{piro2022efficiency}.

One can also consider swimmers with different rigid shapes~\cite{piro2024energetic}, internal degrees of freedom, or shape elasticity.  For example, ``gather-move-spread" is the most efficient strategy to transport a slender, deformable drop of an active fluid along a surface with minimal viscous dissipation~\cite{Shankar-2022-Proc.Natl.Acad.Sci.}. 
Optimal control was also used to maximize the speed of a dung beetle that switches strategies between rolling and reorientation in a noisy environment~\cite{mori2023optimal}

In parallel to the above studies that use optimal control for the navigation problem are a series of studies based on reinforcement learning, reviewed in \cite{nasiri2023optimal}.  These studies allow for unknown stochastic (even turbulent) environments.

\section{CONTROL IN SOFT AND ACTIVE MATTER}
\label{sec:control_soft_active}

Here we briefly review key advances of control theory in soft and active-matter systems, with special emphasis on optimal control methods.
Although optimal control has recently begun to find applications in active condensed matter, applications in passive condensed-matter systems have a long history that extends even before Pontryagin's formulation of the minimum principle.

\subsection{Granular Materials}
\label{sec:granular}
\textit{Granular materials} consist of macroscopic solid particles (or ``grains'') that are typically in a disordered arrangement that may flow.
Interestingly, the field of optimal transport \cite{Villani2003} began with these systems.
In 1781, Gaspard Monge first formalized the optimal-transport problem as one of efficiently transforming any pile of grains into another target pile with minimal work, for the purposes of earthmoving \cite{Monge1781}.
This problem has been further extended to the handling of granular materials in a wide range of applications.
Earlier studies have developed control protocols for automated elements that interact with granular materials, such as shakers, drums, and arms.
For example, optimal protocols have been developed to maximize grain size by controlling moisture content, rotation rates, and bed depths of rotating drums during granulation processes \cite{Wang2006}. 
Similarly, protocols have been developed to optimally control dehydration and granulation of mineral fertilizers in fluidized beds \cite{Korniyenko2024}.
Optimal transport has been employed to achieve target configurations of poured \cite{Tuomainen2022} and swept \cite{alatur2023granular} grains.
Target shapes of deformable foams and systems of confined grains in containers have also been achieved by optimizing robotic-grip configuration sequences \cite{Li2019} and container-shaking protocols \cite{aoyama2024granular}, respectively.

Granular materials continue to attract fundamental interest because the relationship between the structure of the material and its mechanical response under external loading is highly nontrivial and difficult to generalize.
It is a current outstanding goal of the granular-materials community to fully characterize that relationship in order to predict how and where granular materials rearrange and ultimately flow under external loading.
Such predictive capability would ultimately enable the design of mechanical response in this class of materials, with relevance to manufacturing, construction, landscape evolution, and even cancer research.

Optimal control is an attractive route towards understanding and designing granular response to external forcing.
Recent work used optimal control to minimize the connection time between nonequilibrium steady states of a driven granular gas, finding that the optimal (bang-bang) protocol consisted of heating at large driving and cooling at zero driving \cite{prados2021granular}.
A follow-up study found that the optimal connection time depends on the driving intensities \cite{ruiz-pino2022granular}.
Other work has examined force transmission through granular materials, and its relationship to particle rearrangement, through the lens of network optimization.
Force networks within jammed granular solids were investigated as evolving flow networks, and solutions to the maximum-flow minimum-cost problem were associated with locations of failure under compression \cite{Lin2014}.
We believe that there is much opportunity for further employment of control theory to understand the fundamental physics of yield in granular materials.

\subsection{Colloidal Systems}
\label{sec:colloidal}
\textit{Colloidal systems} comprise passive microscopic solid particles immersed in fluids and subject to thermal fluctuations.
Colloidal self-assembly is a well-established and effective means through which nano- and micro-scale structure (and ultimately, materials functionality) can be precisely designed.
The self-assembly process, however, is not always a straightforward march toward lowest free-energy structures: systems can be caught in metastable traps, kinetics play an important role, and environmental conditions often must be carefully and deliberately manipulated to ensure success.

To facilitate (or \emph{direct}) self-assembly, it is common to tune control parameters such as electromagnetic fields, solvent properties, flow dynamics, grafted ligands, temperature, external shearing, and optical tweezers (reviewed in \cite{Liu2023, Solomon2018}).
Optimal control has proven useful in this context, to determine tuning protocols that can drive the dynamical process of self-assembly toward target outcomes (as reviewed in \cite{Paulson2015, Tang2022, N.McDonald-2023-SoftMatter}).
Early work on single particles showed that open-loop protocols could accelerate the relaxation to thermal equilibrium following a translation or a change in confinement (trap strength) via ``engineered swift equilibration"~\cite{martinez2016engineered} or, more generally, ``shortcuts" of various types~\cite{gueryodelin2023driving}.
In many-particle systems, on/off switching of some external parameter has been shown to induce the self-assembly of paramagnetic colloids via a toggled applied magnetic field \cite{Swan2014,Sherman2019}, photo-switchable nanoparticles via pulsed light \cite{Jha2011}, and oppositely charged colloids via oscillating pH levels \cite{Long2018}.
Feedback related to particle bonding reversibility has been incorporated into simulations of sticky-sphere crystallization, in order to develop effective protocols for changing particle interaction strength to improve assembly \cite{Klotsa2013}.
Additionally, it was shown that open-loop control of external charge placement could drive nanoparticle self-assembly into target structures in minimal time \cite{Ramaswamy2015}.
Stochastic dynamic programming was employed to automate particle transport via optical tweezers, in order to avoid collisions in a stochastic environment \cite{Banerjee2010}.
It was also theoretically demonstrated that optimal feedback control of osmotic pressure could drive colloidal depletion and self-assembly into target crystal structures \cite{Xue2013}. 

Machine learning has recently been incorporated into the optimization of colloidal self-assembly.
Operating within the framework of optimal control, reinforcement learning was used to design circular colloidal crystals via an applied electric field \cite{Zhang2020a}.
By contrast, a neural network trained by reinforcement learning was recently used to generate optimal protocols for the self-assembly of patchy colloids, without any reference to formal control theory \cite{Whitelam2020}.
Related work has used automatic differentiation to control crystallization rates \cite{Goodrich2021} and final structures \cite{King2024} by tuning particle interactions.
Recent studies have also used machine learning to perform optimal control. Reinforcement learning can be used to determine a spatiotemporal pattern producing functionality \cite{Falk-2021-Phys.Rev.Res.}.

\subsection{Active Brownian Particles (ABP)}

So far, we have focused on passive systems, whose simplicity readily lends them to theoretical modeling, optimization, and engineering applications.
We now consider active systems.
One paradigmatic example of active matter is suspensions of colloidal particles undergoing Brownian motion while also exhibiting self-propulsion, called \textit{active Brownian particles} (“ABPs”). Example systems include suspensions of Janus particles, as well as collectives of motile organisms such as bacteria and algae.

Experimental methods have been developed to control swimming activity with light.
Light-responsive self-propelled Janus particles were synthesized by selectively coating colloids to expose a patch of surface that reacts with the solvent when illuminated \cite{Palacci-2013-Science}.
This study demonstrated reversible switching between uniform and phase-separated states, confirming predictions of \textit{motility-induced phase separation} \cite{Fily-2012-PhysRevLett}.
Similar methods were extended to design control protocols that allow Janus particles to navigate a maze \cite{Yang-2018-ACSNano}.
Proteorhodopsin, a light-sensitive protein in E.\ coli, was exploited to control the density \cite{Frangipane-2018-eLife, Arlt-2018-NatCommun} and athermal fluctuations \cite{Massana-Cid-2024-NatCommun} of active bacterial suspensions.
In dense algal suspensions, light illumination produces cell aggregation \cite{Dervaux-2017-NaturePhys}.

Feedback control and MPC in ABPs has been extensively reviewed in \cite{Takatori-2025-Annu.Rev.Condens.MatterPhys.}.
Feedback control was applied to dense suspensions of phototactic bacteria to rectify their collective locomotion \cite{Massana-Cid-2022-NatCommun}.
Reinforcement learning has been applied to steering ABPs \cite{Colabrese-2017-Phys.Rev.Lett., Muinos-Landin-2021-Sci.Robot.}.
Closed-loop feedback control has also been used to induce self-assembly of colloidal systems.
For example, closed-loop feedback control of an applied quadrupolar electric field was used to drive the self-assembly of colloidal spheres into two-dimensional crystals, using defect correction in real time \cite{Juarez2012,Tang2016}.
Relatedly, PID control of an applied electric field in a microfluidic device was used to direct colloidal self-assembly \cite{Gao2019a}.
More recently, an analytic solution to control of ABPs was developed, facilitating computation of optimal protocols \cite{Baldovin-2023-Phys.Rev.Lett.}.
An immediate extension of optimal control of ABPs lies in the growing field of \textit{colloidal robotics}~\cite{Liu2023}.
Colloidal robots generally collect local information on the microscale and act in response to that information, using protocols of varying complexity.

\subsection{Active Liquid Crystals and Gels}
\label{sec:activeLCgels}

\textit{Active liquid crystals} represent an additional class of active-matter systems.
These resemble ABP's, except that the constituent particles approximate elongated rods rather than spheres.
The aligned nature of the particles permits a continuum description based on nematic liquid crystals.
Many experimental systems of active liquid crystals are ultimately composed of the cytoskeletal polymers microtubules and actin, though they may be assembled in intracellular structures like cortices and spindles; in supracellular structures like cables; or in reconstituted systems.
These polymers are then driven by their respective associated motor proteins, kinesin/dynein and myosin.
This microscopic driving can be modeled as an active dipolar stress field.
Extensile stresses result in actively stirred liquid crystals, where pairs of oppositely charged nematic defects are created and annihilated \cite{Sanchez-2012-Naturea}.
Although a full description of the material dynamics requires equations of motion for spatially resolved vector and tensor fields, the defects themselves exhibit simpler dynamics akin to self-propelled particles \cite{Giomi-2014-Philos.Trans.R.Soc.Math.Phys.Eng.Sci.}.
As we shall see, this simplification has been recently exploited to facilitate control.

Recent advances in optogenetics have ushered in a series of experimental studies that have begun to establish control over active liquid crystals.
Light-sensitive iLID-kinesin constructs have enabled spatiotemporal control over flows in microtubule-based active liquid crystals \cite{Ross-2019-Nature}.
By determining the appropriate spatiotemporal light input, one can control the microtubule suspensions' patterning \cite{Najma-2022-NatCommun} and flow dynamics \cite{Lemma-2023-PNASNexus}.
Meanwhile, for actin-based active liquid crystals, light-sensitive LOV-myosin-IX was developed and used to demonstrate control over defect dynamics \cite{Zhang-2021-NatMater}.

This degree of experimental control has allowed for sophisticated theoretical studies that seek the spatiotemporal activation sequence necessary to generate desired behavior.
A critical advance demonstrated that activity gradients drive defect dynamics, similarly to electric fields driving charges \cite{Shankar-2019-Phys.Rev.X}.
Subsequently, an optimal control framework was established to determine the perturbation needed to switch between flow directions \cite{Norton-2020-Phys.Rev.Lett.}.
Optimal control has been extended to stabilize otherwise-unstable Couette flows \cite{Ghosh-2024-a}, to control the trajectory of microtubule asters \cite{ghosh_spatiotemporal_2024}, and to control the location of attractors in epithelial tissues \cite{Sinigaglia-2024-Phys.Rev.Lett.}.
More recent work discovered remarkably simple selection rules based on defect symmetry and the spatial profile of activation, which allowed for simple control over defect dynamics \cite{Shankar-2024-Proc.Natl.Acad.Sci.}.
Similarities between the recent development of nematic-defect control and the longstanding problem of navigation (§\ref{sec:swimmers}) could be leveraged in future studies to facilitate control over dynamics.

Active liquid crystals resemble incompressible fluids in their ability to exhibit spontaneous flow.
Introducing connectivity (e.g., in the form of crosslinks) produces passive elasticity and allows motors to cause contraction.
In this limit, active force dipoles are contractile and cause density instabilities in \textit{contractile active gels}. Contractility in reconstituted actomyosin gels has been modulated with light using blebbistatin \cite{Schuppler-2016-NatCommun, Linsmeier-2016-NatCommuna} and caged ATP \cite{Clarke-2025-a}.
Meanwhile, optogenetic control of upstream myosin regulators has been recently developed in starfish oocytes, opening up exciting avenues in controlling contractility \textit{in vivo} \cite{Liu-2025-Nat.Phys.}.
An additional active gel system comprises \textit{pulsatile active gels} of beads that incorporate the oscillatory Belousov-Zhabotinsky reaction.
Optimization was recently employed to maximize their actuation \cite{Blanc-2024-Langmuir}.
Coupling among beads in a collective produces quorum-sensing behavior reminiscent of bacterial populations \cite{Blanc-2024-Proc.Natl.Acad.Sci.}.

Control over active materials has enabled various applications.
Nematic defects can trap colloidal particles, enabling control over their motion and interaction valency \cite{Nelson-2002-NanoLett.}.
Furthermore, control over the resulting hydrodynamic flow fields opens up novel methods in microfluidics \cite{Yang-2025-Nat.Mater.}.
In addition, understanding how nematic defects are controlled has important biological significance.
The nematic actomyosin cortex of the aquatic invertebrate \textit{Hydra} exhibits active defects, which determine the location where morphological features such as limbs regenerate \cite{Maroudas-Sacks-2021-Nat.Phys.}.
Similar behavior occurs in multicellular assemblies.
Epithelial cells tend to extrude near defects \cite{Saw-2017-Naturea}.
In swarms of \textit{Myxococcus} bacteria, cell extrusion at defects nucleates fruiting bodies, which initiate sporulation \cite{Copenhagen-2021-Nat.Phys.}.
It would be interesting to study how biochemical control signals interact with active liquid crystals; a recent theoretical model has begun to investigate this interplay \cite{Norton-2024-Phys.Rev.E}.

 \subsection{Stochastic Systems}

Optimal control has been applied to a variety of \textit{stochastic systems}, where time-dependent variation of the potential energy and nonequilibrium driving forces can drive systems between desired endpoint ensembles.
\textit{Parametric control} involves the dynamic variation of a finite number (typically, only a few) control parameters (thus specifying equilibrium or steady-state distributions), often with the goal of minimizing the required mean work.  
Utilizing the approximation of endoreversibility, the thermodynamic length (a measure of statistical difference) separating equilibrium probability distributions for control-parameter endpoints was related to the minimum dissipated availability during finite-time driving~\cite{salamon_thermodynamic_1983}.
This thermodynamic length was explicitly formulated in stochastic thermodynamics in terms of the Fisher information metric~\cite{crooks_measuring_2007}.
Linear-response theory was used to incorporate dynamics and quantify the dissipated work in terms of a generalized friction metric on the space of control parameters~\cite{sivak_thermodynamic_2012}.
This framework has been used to understand minimum-work control of many model systems~\cite{blaber_optimal_2023} including experimental unfolding of DNA hairpins~\cite{tafoya_using_2019} and driving a model molecular machine~\cite{gupta_optimal_2022}.
For quadratic potentials, the minimum-work control~\cite{schmiedl2007optimal} and resulting system trajectory ensemble obey a surprisingly general time-reversal symmetry~\cite{loos_universal_2024}.

When the controlled system is active, the additional housekeeping energy consumption associated with activity leads generically to a trade-off between the excess costs (higher when driving faster) and the housekeeping costs (higher when driving slower and for longer duration)~\cite{Davis-2024-Phys.Rev.X}.  By contrast, using feedback, for example by incorporating the observed state at the start of the protocol, lowers the dissipation cost~\cite{garcia-millan2024optimal,schuttler2025active}.
An outstanding challenge in stochastic systems is to understand how activation via control signals results in free-energy consumption and work production of actuators, such as molecular motors; to this end, recent studies have developed thermodynamically consistent models of active matter \cite{Markovich-2021-Phys.Rev.Xa, Fodor-2022-Annu.Rev.Condens.MatterPhys.b} and energetic control \cite{Gupta-2023-Phys.Rev.E, Davis-2024-Phys.Rev.X}.
Recent efforts have also applied machine-learning algorithms such as automatic differentiation~\cite{engel_optimal_2022} and genetic algorithms~\cite{whitelam_demon_2023} to learn optimal feedforward and feedback protocols.

\textit{Optimal transport} (see also §\ref{sec:granular}) drives systems between desired endpoint probability distributions, often assuming control over the dynamic variation of the full (continuous-space) potential-energy surface.
Ref.~\cite{aurell_optimal_2011} related optimal transport in a fluid-mechanics framework to minimizing energy expenditure during driven transitions, ushering in the use of optimal transport for nonequilibrium thermodynamic systems, where the $L_2$-Wasserstein distance bounds the entropy production~\cite{nakazato_geometrical_2021}.  
Ref.~\cite{proesmans_finite-time_2020} extended these ideas to the more practically relevant situation of coarsely defined (``mesoscopic'') end states. 
Recent work has uncovered relations between parametric control and optimal transport~\cite{zhong_beyond_2024}.


\subsection{Biological Systems}

Control signals abound in life.
Inside individual cells, signals are sent through small molecules, proteins, signaling cascades, and regulatory networks.
Similarly, multicellular collections communicate by transmitting molecules and mechanical stresses.
Furthermore, neuronal action potentials conduct signals across organismal scales.
These control signals directly influence motor proteins, gene transcription, and ultimately biological function.
Although researchers have developed a library of optogenetic mutants to influence biological systems \cite{Boyden-2005-NatNeurosci}, here we instead review some examples of control-theoretic approaches that aim to understand control strategies in the wild type.

In cells, \textit{regulatory networks} seek to achieve homeostasis yet respond appropriately to varying external signals.
These systems face challenges from fluctuating environments (external noise) and low-copy-number fluctuations of their own chemical species (internal noise)~\cite{ElowitzInternalExternalNoise}.  
Much research has focused on network topologies that can reduce the stochasticity of the internal chemical species~\cite{lestas_fundamental_2010}.
The so-called ``antithetic'' regulatory motif implements integral feedback control of a given chemical species and achieves robust perfect adaptation~\cite{aoki_universal_2019}.
Integral feedback control has proven useful in quantitatively describing chemotaxis in bacteria \cite{Yi-2007-MethodsinEnzymology} and stability to perturbation in animal locomotion \cite{Cowan-2014-IntegrativeandComparativeBiology}

Control theory applied to \textit{complex neural networks} has a rich history outside the scope of this review; recent reviews focused exclusively on network control include \cite{liu2016control, DSouza2023network}, especially in the context of neuroscience \cite{Wu2024}.
Complex neural networks represent the connectivity of brain regions as a set of nodes and edges, with edge weights determined either from physical white-matter connections between regions or correlations of activity patterns across regions \cite{Kulkarni2025}.
Network control theory offers an informative and exciting lens through which state transitions in the brain can be understood and potentially affected via clinical intervention. 
Analysis of human white-matter networks using diffusion spectrum imaging found that brain regions can be categorized according to their controllability, or their role in facilitating theoretical transitions between cognitive states \cite{Gu2015}.
A follow-up study found that average controllability of white-matter networks increases with human age, indicating that white matter develops to optimally support neural dynamics \cite{Tang2017}.
Other studies have also used network control to examine network models of the brain (both human and animal), and gained insights regarding the control architectures of the networks \cite{Betzel2016a, Lee2019a, Luppi2024}, 
the ability of the networks to support different dynamics while maintaining robustness against perturbations \cite{Kim2018a}, and how control mechanisms are impacted by development \cite{Cui2020,Parkes2022}, traumatic brain injury \cite{Gu2017}, schizophrenia \cite{Tang2022a}, and dementia \cite{Meyer-Base2020}.

A paramount problem of control in biology is the self-assembly of complexes, where assemblies of interest are often not ground states and kinetic traps typically complicate ever reaching equilibrium or steady states~\cite{nguyen_organization_2021} (see also §\ref{sec:colloidal}). A model biological self-assembly system is a viral capsid, which has a particular target geometry and many unviable trapped complexes~\cite{HaganARPC}. 
Gradient descent and Markov-state modeling found nonequilibrium protocols that improved yield of target energy minima, metastable states, and transient states for capsid assembly and biopolymer folding~\cite{trubiano_optimization_2022}. Automatic differentiation has also found baseline kinetic rates and time-dependent protocols that significantly increase yield~\cite{JohnsonOptimalKineticPathways}.

\section{OUTLOOK}

In this article, we have provided a brief introduction to control theory and optimal control, aimed at physicists with little prior experience with control theory.
Given the prevalence and longstanding history of control-theoretic methods, we expect to see continued application across a wide range of soft and active condensed-matter systems.
In particular, it would be interesting to further study how control over nonequilibrium activity ultimately underlies biological function.
In biology, physics and control cannot be separated; understanding their interplay is essential toward a holistic description of the physics of life.


\begin{tcolorbox}\textbf{SUMMARY POINTS}
\begin{enumerate}
\item Control theory describes how a control signal (“input”) $u$ influences the dynamics of a system, $\dot x = f(x,u)$.
\item Feedforward control is conceptually simple because it relies on an accurate predictive model of a system's dynamics; however, disturbances and deviations between the model and the actual system (e.g., perturbations and inaccuracies) lead to unpredictable behavior.
\item Feedback control rests on information gained from sensors that report a system's state, allowing the controller to correct for unexpected disturbances; however, multiple feedback gains must be tuned per degree of freedom, usually by trial-and-error.
\item Pontryagin's Minimum Principle (PMP) is conceptually simple; however, a two-point boundary-value problem must be solved for each initial condition.
\item The Hamilton-Jacobi-Bellman equation (HJB) solves for all optimal paths, allowing  optimal control to be robust in the presence of disturbances; however, the large dimensionality makes the partial differential equations typically difficult to solve.
\item The HJB equation can be generalized to systems undergoing stochastic fluctuations, making this approach well-suited to soft and active systems.
\end{enumerate}
\end{tcolorbox}

\begin{tcolorbox}\textbf{FUTURE ISSUES}
\begin{enumerate}
\item We expect that the optimal control methods introduced here will enable more sophisticated control over soft and active systems.
\item To control high-dimensional systems such as collective motion and pattern formation, feedforward optimal control methods such as the Hamiltonian bridge have begun to prove beneficial.
\item Because control signals steer nonequilibrium actuation in living systems, we expect control theory and optimal control to prove essential in understanding the physics of life.
\end{enumerate}
\end{tcolorbox}

\section*{DISCLOSURE STATEMENT}
The authors are not aware of any affiliations, memberships, funding, or financial holdings that might be perceived as affecting the objectivity of this review.

\section*{ACKNOWLEDGMENTS}
We gratefully acknowledge fruitful discussions with Suri Vaikuntanathan, Sarah Loos, Michael Hagan, and Stephen Whitelam; and thank 
Lakshminarayanan 
Mahadavan, Aparna Baskaran, Suraj Shankar, and {\'E}tienne Fodor for feedback on the manuscript. 
JA acknowledges primary support from the National Science Foundation through grant DMR-2144380, additional partial support through the Center for Dynamics and Control of Materials: an NSF MRSEC under Cooperative Agreement No. DMR-2308817, and additional partial support through NSF PHY-2309135 to the Kavli Institute for Theoretical Physics (KITP).
EGT acknowledges support from the International Human Frontier Science Program Organization (HFSPO) under grant RGEC33/2024.
DAS acknowledges a Natural Sciences and Engineering Research Council of Canada (NSERC) Discovery Grant and Discovery Accelerator Supplement RGPIN-2020-04950 and a Tier-II Canada Research Chair CRC-2020-00098.
JB acknowledges an NSERC Discovery Grant.

\newpage

\appendix
\onecolumngrid
\section*{Appendices}
In 
these appendices, 
we illustrate some of the basic techniques for optimal control, in hopes that a simple example solved several ways can illuminate and make concrete the broader discussion given in the main text.   Although our choice is just the ``physicists' favorite toy model," the simple harmonic oscillator, we will find that it is similar to and sheds light on many examples of current interest in soft and active-matter physics.  For more examples, see \cite{bechhoefer2021control,kirk1998optimal}. 

\section{A basic example}

Consider the transport of a particle in a harmonic potential from one position to another in a specified time interval (protocol duration) $\tau$.  The equation of motion and initial and final conditions are
\begin{align}
	\underbrace{\ddot{x}+x = u}_\textrm{dynamics}  \qquad 
	\underbrace{x(0) = x_0 \,, \; \dot{x}(0) = 0}_\textrm{initial}  \qquad 
	\underbrace{x(\tau) = \dot{x}(\tau) = 0}_\textrm{final} \qquad 
	 \,,
\label{eq:equations-2ndOrder}
\end{align}
where $x$ is the one-dimensional position of the particle and $u(t)$ is the applied force at time $t$.  All units are scaled to eliminate explicit parameters such as mass, stiffness, and angular frequency.  The initial condition at $t=0$ corresponds to a stationary particle at position $x_0$, and the final condition at the end of the protocol is at $t=\tau$ and corresponds to a stationary particle at position $x(\tau)=0$.  For now, we neglect damping and stochastic (thermal) forces.

\begin{figure} [th]
\begin{center}
	\includegraphics[width=6.0in]{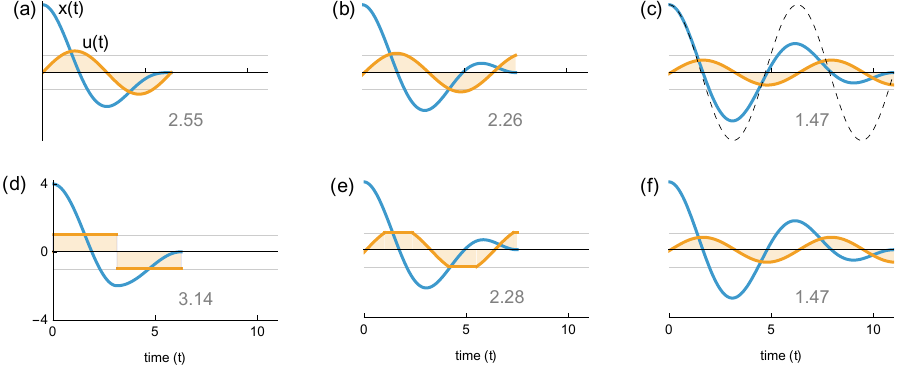}
	\caption{\label{fig:FF}Unconstrained and constrained feedforward control. Protocol duration $\tau$ is (a,d) $2\pi$, (b,e) 7.5, or (c,f) 11.  Blue curve is the particle position $x_1^*(t)$.  Orange-brown filled curve is the control input $u^*(t)$.  Initial position is $x_0 = 4$.  Control costs $J$ for each case are in gray at bottom center-right.  (a--c):  Unconstrained input.  (d--f):  Input constrained to $|u(t)| \le 1$. The black dashed curve in (c) shows the uncontrolled sinusoidal oscillations.}
\end{center}
\end{figure}

We seek the control $u$ that moves the particle from $x_0$ to the origin, in time $\tau$, while minimizing the total cost
\begin{align}
	J =  \frac{1}{2} \int_0^\tau \dd{t} u^2(t) \,.
\label{eq:cost}
\end{align}
Here, $J$ is the integral of the \textit{running cost} $\tfrac{1}{2} u^2$.  Here, the cost depends only on the input $u$ and implies moving from $x_0$ to $0$ in time $\tau$ while using a minimum of ``control effort."  Absent any control, the particle will oscillate forever about the desired position, $x=0$
(Fig.~\ref{fig:FF}c.)  In Sec.~\ref{sec:relations2stochastic}, we discuss the relation between effort and physical quantities such as energy.

The control can be given either as a \textit{feedforward command} $u(t)$ or as a \textit{feedback command} $u(\bm{x}(t),t)$ that also depends on the instantaneous particle state. Here, $\bm{x}$ denotes the two-dimensional state vector formed from $x_1=x$ and $x_2=\dot{x}$.  In vector-matrix notation, the equations of motion are  $\dot{\bm{x}} = \bm{Ax} + \bm{B}u$.  More explicitly,  \begin{align}
	\dv{\bm{x}}{t} = \dv{t} \mqty(x_1 \\ x_2) = \underbrace{\mqty(0 & 1 \\ -1 & 0)}_{\bm{A}}  
		\mqty(x_1 \\ x_2) + \underbrace{\mqty(0 \\ 1)}_{\bm{B}} u \,,
\label{eq:linear}
\end{align}
with $\bm{A}$ the dynamics matrix and $\bm{B}$ the input coupling.  The input $u$ (\textit{control parameter}, in the physics literature) here is a scalar because there is only one input function, but more generally would be a vector. 

Note that in traditional problems of classical mechanics, we specify the initial conditions and input $u(t)$ and then determine the trajectory $\bm{x}(t)$ and, with it, the final condition $\bm{x}(\tau)$.  Here, instead of an \textit{initial value problem}, we have a \textit{boundary value problem}: we specify $\bm{x}$ at two different times but seek a $u(t)$ that connects them.  The goal for optimal control is to choose the $u(t)=u^*(t)$ from the infinity of possible solutions so that $u^*(t)$ minimizes the cost functional $J$ of Eq.~\eqref{eq:cost}.

A helpful check and aid in solving problems is that when $H$ has no explicit time dependence (i.e., when both the running cost and the equations of motion are time invariant), the Hamiltonian is constant when evaluated on optimal trajectories $(\bm{x}^*(t), \bm{\lambda}^*(t),u^*(t))$.  For smooth solutions, the argument is the same as used in classical mechanics to identify a Hamiltonian with the total energy:  $\dot{H} = \partial_x H \dot{x} + \partial_\lambda H \dot{\lambda} + \partial_u H \dot{u} = 0$, after substituting Hamilton's equations.

\section{Unconstrained control:  Feedforward solution}
\label{sec:unconstrainedFeedforward}
If there are no constraints on the magnitude of the allowed control $u(t)$, then we can solve the optimal-control problem by the calculus of variations to find the optimal feedforward control $u(t)$.  As in the main text, the Hamiltonian form is convenient.  The control Hamiltonian $H$ is
\begin{align}
	H(\bm{x},\bm{\lambda},u) 
		= \frac{1}{2}u^2 + \lambda_1 x_2 + \lambda_2(-x_1 + u) \,.
\label{eq:hamiltonian}
\end{align}
The equation $\dot{\bm{x}} = \partial_{\bm{\lambda}} H$ reproduces Eq.~\eqref{eq:linear}.  The adjoint equation $\dot{\bm{\lambda}} = -\partial_{\bm{x}} H$ gives $\dot{\lambda_1}=\lambda_2$ and $\dot{\lambda}_2 = - \lambda_1$.  The compatibility equation $\partial_u H = 0$ leads to $u = -\lambda_2$.  Substituting for $u$ then gives coupled linear differential equations for the enlarged state $(x_1,x_2,\lambda_1,\lambda_2)^\tran$ with the four boundary conditions $x_1(0)=x_0$ and $x_2(0) = x_1(\tau) = x_2(\tau) = 0$.

Figure~\ref{fig:FF} shows three examples of an unconstrained protocol and the resulting particle position, for $x_0=4$ and $\tau = 2\pi$, 7.5, and 11.  
Notice that in all three cases, the derivative is flat (the particle is stationary) at $t=0$ and $\tau$.  It is easy to confirm that the control cost (shown in gray at bottom center of each plot) decreases asymptotically as $1/\tau$~\cite{sekimoto1997complementarity,schmiedl2007optimal}.  Notice that the control parameter (force) is not applied monotonically; rather, it mostly brakes the oscillatory motion, providing a controlled feedback damping for the otherwise-dissipationless system.

\section{Constrained control}
\label{sec:constrained}

In a physical experiment, there will always be limits on the control input $u(t)$.  The Pontryagin Minimum (Maximum) Principle (PMP) gives a systematic way to derive minimum-cost solutions in the presence of such constraints.  As explained in the main text, 
if the value of $u(t)$ minimizing $H$ lies within the allowed set, it can be determined from the compatibility condition using $\partial_u H(\bm{x},\bm{\lambda},u) = 0$.  If not, one should check the boundary values of $u$ and see which minimizes $H$.  Here, we let the set of allowable (scaled) forces be $|u| \le 1$ and check the cases $u = \pm 1$.  Since $u^2=1$ in this case, the only relevant term in $H$ is $\lambda_2 u$, implying $u = -$sign$(\lambda_2)$.  Putting together the ``interior'' and boundary solutions, $u^*(t) = -$sat$(\lambda_2(t))$ is valid for all values of $\lambda_2$, where sat$(\cdot)$ clips its argument whenever its magnitude exceeds 1.  Thus, in this case, the only change in the coupled state-adjoint system of equations is that the equation $\dot{x}_2 = -x_1 - \lambda_2$ becomes $\dot{x}_2 = -x_1 -$ sat$(\lambda_2)$.  

In Fig.~\ref{fig:FF}(d--f), we plot the constrained solutions corresponding to the protocol durations used in (a--c), where there are three cases.  For large $\tau$ (Case f), the solution is continuous and identical to the uncontrolled solution.  Indeed, in this problem, as the protocol duration is extended, solutions with decreasing amplitude for $u(t)$ become possible.  The amplitude scales as $u \sim 1/\tau$, meaning the cost scales as 
\begin{align}
	J \sim \int_0^\tau \dd{t} u^2 \sim (\tau)(1/\tau^2) = 1/\tau \,,
\end{align}	
as mentioned at the end of Sec.~\ref{sec:unconstrainedFeedforward}. The costs $J$ for the constrained solutions are equal or greater than the costs of the unconstrained solutions.

Figure~\ref{fig:FF}d is an example of \textit{bang-bang control}, since $u(t)$ alternates exclusively between its limits $\pm 1$.  The protocol duration $\tau=2\pi$ was chosen to be the minimum at which there is a solution meeting the boundary conditions at $t=0$ and $\tau$.  This could be found numerically by computing the solutions for different values of $\tau$; alternatively, in Sec.~\ref{sec:minimumTime} we use the PMP to analytically find the minimum duration.

The last case, Fig.~\ref{fig:FF}e, shows a complicated mixed solution, with repeated alternations between continuously varying solutions from the interior problem and boundary solutions.

A feature common to all the solutions is that there are discontinuities in either $u(t)$ or its temporal derivative $\dot{u}(t)$.  These are typical in optimal-control problems, and it is worth understanding their origin and implications.  We begin by noting that even the unconstrained solutions in Fig.~\ref{fig:FF}(a--c) can show discontinuities at the initial or final times of the protocol.  These are induced by the particular choice of boundary conditions.  For example, at the end, we require that $x(\tau) = \dot{x}(\tau) = 0$.  In a slightly modified problem, we might be satisfied if at $t=\tau$ the particle is \textit{near} the origin and not moving too quickly, achievable via explicit end-time penalties.  Alternatively, if we add a term $\sim x_1^2(t)$ to the running cost $L$, the final conditions become $\lambda_1(\tau) = \lambda_2(\tau) = 0$.  This follows easily from an analysis using the calculus of variations for an augmented cost that adds the equations of motion as a constraint imposed by a Lagrange multiplier $\bm{\lambda}(t)$.  More intuitively, for finite cost of ending up near (but not exactly at) the desired final state, it is better to shut off the control at the end, setting $u(\tau) = 0$:  In a system with inertia, the benefit of a control applied at time $t$ is only accrued at a later time, whereas the cost is applied at the current time.  The exception is the final condition, which implies that the cost for violating a final condition is infinite.  The control then seeks to satisfy the condition, whatever the future costs.

The other discontinuities present in $u(t)$ result from switching among interior and exterior solutions.  Bang-bang control results from switching suddenly from one extreme to another.  The slope discontinuities arise when passing from exterior to interior solutions for $u$, or vice versa.

Note that there is nothing physically implausible about a control that changes discontinuously.  We expect that physical system states should change continuously, but the control we desire can jump ``instantaneously."  Of course, the control is implemented by a physical system that cannot change instantaneously, but often the times involved are sufficiently short to be negligible.  For example, light intensity, which can affect optimal trapping strength or the rate of chemical reactions, can easily be varied at nanosecond to microsecond time scales.  Those scales are often much faster than those of the physical system under control, and their effects can be neglected.  If not, one can extend the physical model to include the actuator dynamics.  Still, the input to the augmented system could change instantaneously.

Finally, we can confirm numerically that both the constrained \textit{and} unconstrained solutions shown in Fig.~\ref{fig:FF} have Hamiltonians that are constant when evaluated on the solutions.  That is, even for bang-bang solutions where $u(t)$ has jump discontinuities and the simple calculus-of-variations argument given above breaks down, $H$ nonetheless remains constant.\footnote{More precisely, if $u^*(t)$ has a jump discontinuity at $t=t_{\rm jump}$, then $H(t \to t_{\rm jump}^-) = H(t \to t_{\rm jump}^+)$.}

\section{Minimum-time control}
\label{sec:minimumTime}

\begin{figure} [th]
\begin{center}
	\includegraphics[width=2.5in]{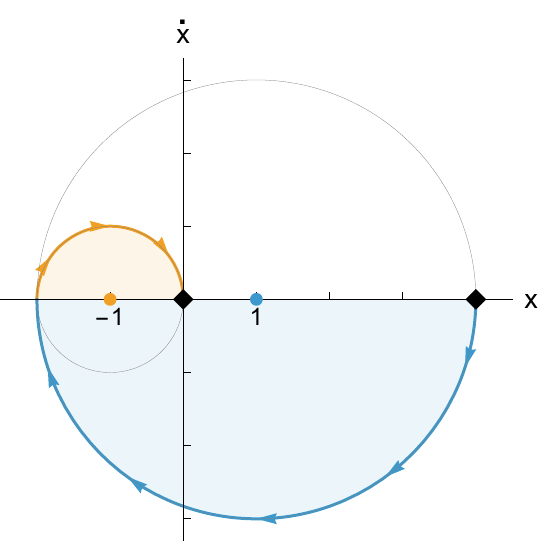}
	\caption{\label{fig:BangBang}Phase-space plot of the bang-bang solution plotted in Fig.~\ref{fig:FF}d.  Black diamonds denote the start and end states. The first stage (blue), from $t=0$ to $\pi$ is a circle of radius 3, centered on $x=+1$.  The second stage (orange), from $t=\pi$ to $2\pi$, is a circle of radius 1, centered on $x=-1$.}
\end{center}
\end{figure}

A natural application of optimal control is to find how to accomplish a task as fast as possible.  For linear dynamics with additive control, such problems have bang-bang solutions set by the actuator limits.  In the example discussed here, we find the minimum-time solution by changing the running cost from $\tfrac{1}{2} u^2$ to 1, meaning that the cost $J=\tau$.  The control Hamiltonian becomes
\begin{align}
	H(\bm{x},\bm{\lambda},u) 
		= \tau + \lambda_1 x_2 + \lambda_2(-x_1 + u) \,.
\label{eq:hamiltonianBB}
\end{align}
Since $H$ is now linear in $u$, the solution always has the bang-bang form.  Proceeding as with the original problem (but reverting to second-order notation for simplicity), we solve the system of equations
\begin{align}
	\ddot{x} + x = -\textrm{sign} (\dot{\lambda}) , \qquad \ddot{\lambda} + \lambda = 0 \,,
\label{eq:BB}
\end{align}
with boundary conditions $x(0) = x_0$, $\dot{x}(0) = x(\tau) = \dot{x}(\tau) = 0$.  Here, $\tau$ is unknown, and we seek the smallest $\tau$ for which a solution exists.  It is easy to verify that, for $x_0=4$, the solution is $\tau=2\pi$.  The control $u^*(t) = -1$ up to $t=\pi$ and $+1$ between $t=\pi$ and $2\pi$. The particle trajectory
\begin{align}
	x^*(t) = \begin{cases}
		1 + 3 \cos t \,, & 0 \le t < \pi \,, \\
		- 1 + \cos t \,,  & \pi \le t \le 2\pi \,,
	\end{cases}
\label{eq:BBsoln}
\end{align}
matches the one found numerically in Fig.~\ref{fig:FF}d.  Notice that the value of the optimal cost found numerically, $\approx 3.14$, is consistent with $J^*=\int_0^{2\pi} \dd{t} \tfrac{1}{2} (\mp 1)^2 = \pi$.  The form $x^*(t)$ of the bang-bang state trajectory is pleasing when plotted in phase space: the motion consists of circular arcs, with the first half centered on $+1$ and the second on $-1$, joining at $t=\pi$ (Fig.~\ref{fig:BangBang}).

\section{Optimal control:  Feedback form}

The above discussion illustrated how one can find feedforward solutions $u^*(t)$ to optimal-control problems.  As discussed in the main text, 
the Hamilton-Jacobi-Bellman equation in principle leads to feedback solutions of the form $u^*(x(t),t)$.  For linear dynamics, the solution can be carried through (semi-) analytically and reproduces the feedforward solution. For simplicity, we assume a continuous, interior solution, but the generalization to constrained control follows a path similar to that taken for the feedforward case.

It will be just as easy to do the derivation for general linear dynamics $\bm{\dot{x}} = \bm{Ax} + \bm{Bu}$ and general quadratic running cost $L(\bm{x},\bm{u}) = \tfrac{1}{2}(\bm{x}^\tran \bm{Qx} + \bm{u}^\tran \bm{Ru})$.  From the main text, 
the HJB equation for linear dynamics is
\begin{align}
	\inf_{\bm{u}} \left[ 
	L(\bm{x},\bm{u}) + 
	\left( \partial_{\bm{x}} J^* \right) \, \bm{f}(\bm{x},\bm{u}) \right] = 0 \,,
\label{eq:HJBlinear}
\end{align}
where $J^*(\bm{x},t)$ is the optimal cost-to-go function starting from $\bm{x}$ at time $t$.  Then,
\begin{align}
	\inf_{\bm{u}} \left[ 
	 \tfrac{1}{2}(\bm{x}^\tran \bm{Qx} + \bm{u}^\tran \bm{Ru}) +
	\left( \partial_{\bm{x}} J^* \right) \,  (\bm{Ax} + \bm{Bu}) \right] = -\partial_t J^* \,.
\end{align}
The suggested \textit{ansatz} $J^* = \tfrac{1}{2} \bm{x}^\tran \bm{S}(t)\bm{x}$ implies that $\bm{S}$ is symmetric, since any antisymmetric component will contribute 0 to $J^*$.  Thus, 
\begin{align}
	\left( \partial_{\bm{x}} J^* \right) = \bm{x}^\tran \bm{S} \,.
\end{align}
Hence, 
\begin{align}
	\inf_{\bm{u}} \left[ 
	 \tfrac{1}{2}(\bm{x}^\tran \bm{Qx} + \bm{u}^\tran \bm{Ru}) +
	\left( \bm{x}^\tran \bm{S} \right) \,  (\bm{Ax} + \bm{Bu}) \right] 
		= -\tfrac{1}{2}  \bm{x}^\tran \dot{\bm{S}} \bm{x} \,.
\end{align}
Because $\bm{u}$ is unbounded, the infimum is found by taking $\partial_{\bm{u}}$ and setting to zero:
\begin{align}
	\bm{u}^\tran \bm{R} + \bm{x}^\tran \bm{S} \, \bm{B}  = \bm{0}^\tran \,.
\end{align}
Taking a transpose and remembering that $\bm{R}$ and $\bm{S}$ are symmetric gives
\begin{align}
	\bm{u} =  -(\bm{R}^{-1} \bm{B}^\tran \bm{S}) \bm{x} \equiv -\bm{K}^\tran \bm{x} \,.
\label{eq:ufb}
\end{align}
Substituting $\bm{u}$ back into the HJB, Eq.~\eqref{eq:HJBlinear}, gives
\begin{subequations}
\begin{align}
	 \tfrac{1}{2} \left[ \bm{x}^\tran \bm{Qx} + (\bm{x}^\tran \bm{SBR}^{-1}) \bm{R}
	 	 (\bm{R}^{-1} \bm{B}^\tran \bm{S} \bm{x}) \right] +
	\left( \bm{x}^\tran \bm{S} \right) \,  \left[ \bm{Ax} 
		- \bm{B}(\bm{R}^{-1} \bm{B}^\tran \bm{S}) \bm{x} \right] 
		&= -\tfrac{1}{2}  \bm{x}^\tran \dot{\bm{S}} \bm{x}  \\
	 \bm{x}^\tran \left[ \tfrac{1}{2} 
	 	\left( \bm{Q} + \bm{SBR}^{-1}\bm{B}^\tran \bm{S} \right)
		+\tfrac{1}{2} \left( \bm{SA} + \bm{A}^\tran \bm{S} \right) 
		- \bm{SBR}^{-1}\bm{B}^\tran \bm{S}\right] \bm{x}	\
		&=-\tfrac{1}{2}  \bm{x}^\tran \dot{\bm{S}} \bm{x}  \,,
\end{align}
\end{subequations}
which implies that $\bm{S}$ obeys the (matrix) Riccati equation,
\begin{align}
	 \bm{Q} + \bm{SA} + \bm{A}^\tran \bm{S} 
	 	- \bm{SBR}^{-1} \bm{B}^\tran \bm{S} = - \dot{\bm{S}} \,.
\label{eq:riccati}
\end{align}
Note the decomposition $\bm{SA} \to \tfrac{1}{2}( \bm{SA} + \bm{A}^\tran \bm{S})$, which follows because the condition $\bm{x}^\tran [\cdots] \bm{x} = 0$ implies that the \textit{symmetric} part of the bracketed terms $[ \cdots ]$ equals zero.  The linear combination isolates the symmetric part of $\bm{SA}$.  There is no constraint placed on antisymmetric terms.

In the problem here, the dynamics is $\bm{A} = \smqty (0 & 1 \\ -1 & 0)$, the input coupling is $\bm{B} = \smqty (0 \\ 1)$, the cost factors are $\bm{Q} = \bm{0}$, and $\bm{R} =1$.  The Riccati equation \eqref{eq:riccati} then reduces to three equations for the components of the symmetric matrix $\bm{S}$:
\begin{subequations}
\begin{align}
	\dot{s}_{11} &= 2s_{12} + s_{12}^2 \\
    \dot{s}_{22} &= -2s_{12} + s_{22}^2 \\
	\dot{s}_{12} &= -s_{11} + s_{22} + s_{12}s_{22} \,,
\label{eq:Seqs}
\end{align}
\end{subequations}
with final conditions $s_{11}(\tau) = s_{22}(\tau) = \infty$ and $s_{12}(\tau) = 0$.  The infinite costs arise because of the final conditions placed on $\bm{x}(\tau)$; if deviations from the desired state instead have finite costs, the final condition would be $\bm{S}(\tau) = \bm{0}$.  Solving these coupled nonlinear equations (the \textit{Matrix Riccati equation}) numerically and setting $u^* = -\bm{K}^\tran \bm{x}$ with $\bm{K}^\tran = \bm{B}^\tran \bm{S}$, we find precisely the different unconstrained $u^*$ shown in Fig.~\ref{fig:FF}a--c,f. 

\begin{figure} [h]
\begin{center}
	\includegraphics[width=6.0in]{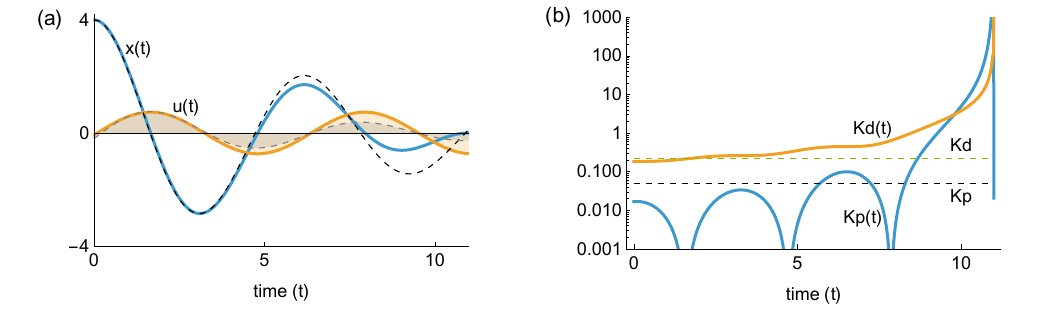}
	\caption{\label{fig:FB}Comparison of optimal and heuristic feedback control.  (a) Solid blue and orange curves replicate using feedback the feedforward curves of Fig.~\ref{fig:FF}c, with $x_0=4$ and $\tau=11$. The dashed curves use proportional-derivative (PD) control.  (b)  Feedback gains for optimal and heuristic control.  The time-dependent solid curves result from the optimal-control calculation, Eqs.~\eqref{eq:ufb} and \eqref{eq:riccati}. 
    The horizontal dashed lines are empirically tuned, constant PD gains $K_{\rm p}$ and $K_{\rm d}$.}
\end{center}
\end{figure}

\section{Heuristic (PD) feedback}
\label{sec:heuristic}

The form of the feedback solution found above is that of a time-dependent linear feedback for the two-component vector $\bm{K}^\tran \equiv (K_p \; K_d)$.  Writing out the feedback law in components and reverting to $x_1 \to x$ and $x_2 \to \dot{x}$ gives 
\begin{align}
	u(x,\dot{x},t) = -K_p(t) x - K_d(t) \dot{x} \,.
\label{eq:PIDtimedep}
\end{align}
Equation~\eqref{eq:PIDtimedep} has the form of proportional-derivative (PD) feedback, one of the heuristic algorithms discussed in the main text.  
The important difference is that the heuristic PD control has constant gains $K_p$ and $K_d$; by contrast, the optimal control has time-dependent gains.

Figure~\ref{fig:FB} compares the optimal and heuristic feedback control algorithms on our example oscillator, for $x_0=4$ and $\tau = 11$.  The solid curves in Fig.~\ref{fig:FB}a reproduce the results from Fig.~\ref{fig:FF}c,f, but they were calculated in a completely different way, from the Riccati equation~\eqref{eq:Seqs}.  Figure~\ref{fig:FB}b shows the time-dependent gains $K_p(t)$ and $K_d(t)$.  Note how they diverge at the end of the protocol, when $t \to \tau$.  

To understand the divergence, Fig.~\ref{fig:FB}b also shows the result of the heuristic PD control, using constant gains $K_p$ and $K_d$ tuned so that the initial decay of $x(t)$ follows the optimal-control solution.  The agreement is reasonable until $t \approx 4$ and then diverges increasingly.  The action of the PD control is easy to understand, if we substitute the feedback $u = -K_p x - K_d \dot{x}$ into the equations of motion, $\ddot{x}+x=u$.  The closed-loop equations are
\begin{align}
	\ddot{x} + K_d \dot{x} + (1+K_p)x = 0 \,.
\end{align}
Thus, proportional gain speeds up the natural frequency by a factor $\sqrt{1+K_p}$, while the derivative gain introduces damping.  The result is a damped harmonic oscillator, whose amplitude oscillates and decays, within an envelope $e^{-(K_d/2)t}$.  However, to stop at finite time (reach $x=\dot{x} = 0$ at time $\tau$), the gain  must diverge: a constant-gain control cannot stop a particle in finite time.  If the protocol time is long, the exponential decay means that there is little difference between the protocols, but at shorter times there is a large difference.

Returning to optimal control, one might wonder, If the feedback and feedforward solutions are identical, why seek feedback solutions?  Given perfect knowledge of dynamics and absent unexpected disturbances, there would be no difference; however, feedback can deal much more robustly with these uncertainties.  Indeed, the best control strategies typically combine feedforward and feedback: feedforward uses the model to implement the optimal control; the feedback around the feedforward solution then deals with the unknowns of modeling errors in the dynamics and external disturbances.

\section{Relations to stochastic, soft, and active-matter problems}
\label{sec:relations2stochastic}

Above, we chose the problem of moving a harmonic oscillator by an applied force because it was a simple physical setting that can be intuitively understood.  This setting is very close to the experiment of Le Cunuder, et al., who studied experimentally an atomic force microscope cantilever whose end position is controlled by forces generated by an external electric potential~\cite{lecunuder2016fast}.  Compared to our toy problem here, the experimental system is subject both to a linear damping term $\gamma \dot{x}$ and to thermal fluctuations, modeled as white noise $\eta(t)$ that has zero mean and variance $\langle \eta(t) \eta(t') \rangle = 2\gamma k_{\rm B}T \delta(t-t')$.  Here $\gamma$ is the linear damping coefficient, $T$ is the environment temperature, and $k_{\rm B}$ is the Boltzmann constant.  The authors did not attempt to minimize the (average) work required to move the cantilever but instead used inverse engineering to find a protocol $u(t)$ that moved the system from an initial equilibrium state at $x_0$ to a final equilibrium state at $x_\tau$.  Because the equations are linear and the noise Gaussian, the distributions of all quantities remain Gaussian and are thus specified entirely by their mean and variance.  Thus, relative to the problem considered above in this Supplement, one needs to match not only the mean positions and velocities at start and end but also the initial and final variances of these two quantities.  The authors chose a polynomial whose order matched the number of boundary-condition constraints.  

Second-order linear systems have also been used to model several other experimental situations.  For example, Loos, et al.\ examined, theoretically and in experiment, the transport of a colloidal particle by a moving harmonic potential in a viscoelastic medium modeled as a Maxwell fluid with a single time constant~\cite{loos_universal_2024}.  The system was then described by
\begin{subequations}
\begin{align}
	\tau_p \dot{x}_p &= -\frac{\kappa}{\kappa_b}(x_p+u) -(x_p-x_b) + \xi_p \\
	\tau_b \dot{x}_b &= -(x_b - x_p) + \xi_b \,,
\end{align}
\end{subequations}
where quantities subscripted by $p$ denote the colloidal particle and quantities denoted by $b$ denote a fictitious ``bath particle" that models the viscoelastic response of the fluid medium.  The dynamics for $x_p$ are that of an overdamped particle, where inertial effects relax so quickly that they may be neglected.  The bath is modeled as another first-order system.  Together, the two form a second-order system of equations with two independent noise sources.  Similar equations also describe active-matter models such as Active Ornstein-Uhlenbeck Particles (AOUP), where a colloidal particle moves in a simple fluid that is nonetheless subject to active fluctuations in the bath~\cite{szamel2014self-propelled,martin2021statistical,saha2023information}. 

One other generalization that arises when we consider soft-matter problems is that the cost function $J$ may have a more complicated form than considered above.  For example, let us consider a colloidal particle moving in a harmonic potential in one dimension.  For the deterministic problem, we used $\tfrac{1}{2} u^2(t)$ as an instantaneous control ``cost" that was to be minimized.  For the stochastic problem, the mean work required to move a harmonic oscillator in a time $\tau$ is given  by~\cite{gomezMarin2008optimal}
\begin{align}
	W = \int_0^\tau \dd{t} \dot{u}(u-\langle x \rangle) \,,
\end{align}
where $\langle x \rangle$ is the time-dependent mean position of the particle, averaged over an ensemble of systems in a fluctuating thermal bath.  The expression for the mean work $W$ can be manipulated to be closer to our $\int \dd{t} \tfrac{1}{2}u^2$ using integration by parts.  The result includes boundary terms (``terminal costs") and quadratic combinations of $u$ and $\langle x \rangle$.  Thus, the cost function falls into the same class as those considered here.


%
%


\bibliography{refs}

\begin{thebibliography}{157}
\expandafter\ifx\csname natexlab\endcsname\relax\def\natexlab#1{#1}\fi

\bibitem{Liu2023}
Liu AT, Hempel M, Yang JF, Brooks AM, Pervan A, et~al. 2023.
\textit{Nature Materials} 22(12):1453--1462

\bibitem{Solomon2018}
Solomon MJ. 2018.
\textit{Langmuir} 34(38):11205--11219

\bibitem{Takatori-2025-Annu.Rev.Condens.MatterPhys.}
Takatori SC, Quah T, Rawlings JB. 2025.
\textit{Annual Review of Condensed Matter Physics} 16:319--341

\bibitem{Paulson2015}
Paulson JA, Mesbah A, Zhu X, Molaro MC, Braatz RD. 2015.
\textit{Journal of Process Control} 27:38--49

\bibitem{Tang2022}
Tang X, Grover MA. 2022.
\textit{Annual Review of Control, Robotics, and Autonomous Systems}
  5(1):491--514

\bibitem{N.McDonald-2023-SoftMatter}
McDonald MN, Zhu Q, Paxton WF, Peterson CK, Tree DR. 2023.
\textit{Soft Matter} 19(9):1675--1694

\bibitem{bechhoefer2021control}
Bechhoefer J. 2021.
Control theory for physicists.
Cambridge Univ. Press

\bibitem{parrondo2015thermodynamics}
Parrondo JMR, Horowitz JM, Sagawa T. 2015.
\textit{Nature Physics} 11:131--139

\bibitem{fliess1995flatness}
Fliess M, L{\'e}vine J, Martin P, Rouchon P. 1995.
\textit{International Journal of Control} 61:1327--1361

\bibitem{mellinger2011minimum}
Mellinger D, Kumar V. 2011.
\textit{Proceedings of the IEEE International Conference on Robotics and
  Automation} :2520--2525

\bibitem{jacobs1894the}
Jacobs J. 1894.
The fables of {Aesop}, chap. {The Old Man and Death}.
Macmillan and Co.,  164--167

\bibitem{ross2012review}
Ross IM, Karpenko M. 2012.
\textit{Annual Reviews in Control} 36:182--197

\bibitem{itzkovitz2012optimality}
Itzkovitz S, Blat IC, Jacks T, Clevers H, {van Oudenaarden} A. 2012.
\textit{Cell} 148:608--619

\bibitem{michaels2019optimal}
Michaels TCT, Weber CA, Mahadevan L. 2019.
\textit{Proceedings of the National Academy of Sciences of the United States of
  America} 116:14593--14598

\bibitem{dear2021feedback}
Dear AJ, Michaels TCT, Knowles TPJ, Mahadevan L. 2021.
\textit{Journal of Chemical Physics} 155:064102

\bibitem{yi2000robust}
Yi TM, Huang Y, Simon MI, Doyle J. 2000.
\textit{Proceedings of the National Academy of Sciences of the United States of
  America} 97:4649--4653

\bibitem{bellman1957dynamic}
Bellman R. 1957.
Dynamic programming.
Princeton Univ. Press

\bibitem{bertsekas2017dynamic}
Bertsekas DP. 2017.
Dynamic programming and optimal control.
vol.~1.
Athena Scientific, 4th ed.

\bibitem{kappen2005path}
Kappen HJ. 2005.
\textit{Journal of Statistical Mechanics} :P11011

\bibitem{kalman1960new}
Kalman RE. 1960.
\textit{ASME Journal of Basic Engineering} 82:35--45

\bibitem{sinha2023optimal}
Sinha S, Krishnan V, Mahadevan L. 2023.
\textit{arXiv:2311.17039}

\bibitem{krishnan2024hamiltonian}
Krishnan V, Sinha S, Mahadevan L. 2024.
\textit{arXiv:2410.12665}

\bibitem{monaghan2012smoothed}
Monaghan J. 2012.
\textit{Annual Review of Fluid Mechanics} 44:323--346

\bibitem{rawlings2024model}
Rawlings JB, Mayne DQ, Diehl MM. 2024.
Model predictive control: Theory, computation, and design.
Nob Hill, 5th ed.

\bibitem{sutton2018reinforcement}
Sutton RS, Barto AG. 2018.
Reinforcement learning: An introduction.
MIT Press, 2nd ed.

\bibitem{silver2018general}
Silver D, Hubert T, Schrittwieser J, Antonoglou I, Lai M, et~al. 2018.
\textit{Science} 362:1140--1144

\bibitem{recht2019tour}
Recht B. 2019.
\textit{Annual Review of Control, Robotics, and Autonomous Systems} 2:253--279

\bibitem{brunton2022data}
Brunton SL, Kutz JN. 2022.
Data driven science and engineering: Machine learning, dynamical systems, and
  control.
Cambridge Univ. Press, 2nd ed.

\bibitem{kaufmann2023champion}
Kaufmann E, Bauersfeld L, Loquercio A, M{\"u}ller M, Koltun V, Scaramuzza D.
  2023.
\textit{Nature} 620

\bibitem{song2023reaching}
Song Y, Romero A, M{\"u}ller M, Koltun V, Scaramuzza D. 2023.
\textit{Science Robotics} 8:eadg1462

\bibitem{sutton1992reinforcement}
Sutton RS, Barto AG, Williams RJ. 1992.
\textit{IEEE Control Systems Magazine} 12:19--22

\bibitem{schmiedl2007optimal}
Schmiedl T, Seifert U. 2007.
\textit{Physical Review Letters} 98:108301

\bibitem{zermelo1931uber}
Zermelo E. 1931.
\textit{Zeitschrift f{\"u}r Angewandte Mathematik und Mechanik} 11:114--124

\bibitem{liebchen2019optimal}
Liebchen B, L{\"o}wen H. 2019.
\textit{Europhysics Letters} 127:34003

\bibitem{daddi2021hydrodynamics}
Daddi-Moussa-Ider A, L{\"o}wen H, Liebchen B. 2021.
\textit{Communications Physics} 4:15

\bibitem{piro2022optimal}
Piro L, Mahault B, Golestanian R. 2022.
\textit{New Journal of Physics} 24:093037

\bibitem{piro2021optimal}
Piro L, Tang E, Golestanian R. 2021.
\textit{Physical Review Research} 3:023125

\bibitem{piro2022efficiency}
Piro L, Golestanian R, Mahault B. 2022.
\textit{Frontiers in Physics} 10:10:1034267

\bibitem{piro2024energetic}
Piro L, Vilfan A, Golestanian R, Mahault B. 2024.
\textit{Physical Review Research} 6:013274

\bibitem{Shankar-2022-Proc.Natl.Acad.Sci.}
Shankar S, Raju V, Mahadevan L. 2022.
\textit{Proceedings of the National Academy of Sciences of the United States of
  America} 119(35):e2121985119

\bibitem{mori2023optimal}
Mori F, Mahadevan L. 2023.
\textit{arXiv:2311.18813}

\bibitem{nasiri2023optimal}
Nasiri M, L{\"o}wen H, Liebchen B. 2023.
\textit{Europhysics Letters} 142:17001

\bibitem{Villani2003}
Villani C. 2003.
Topics in optimal transportation.
Graduate Studies in Mathematics. American Mathematical Society

\bibitem{Monge1781}
Monge G. 1781.
Mémoire sur la théorie des déblais et des remblais.
Histoire de l’{{Academie Royale}} Des {{Sciences}} de {{Paris}}. Imprimerie
  royale

\bibitem{Wang2006}
Wang FY, Ge XY, Balliu N, Cameron IT. 2006.
\textit{Chemical Engineering Science} 61(1):257--267

\bibitem{Korniyenko2024}
Korniyenko BY, Ladieva LR, Pisarenko VG, Pisarenko JV, Nesteruk AO. 2024.
\textit{Cybernetics and Systems Analysis} 60(5):726--735

\bibitem{Tuomainen2022}
Tuomainen N, {Blanco-Mulero} D, Kyrki V. 2022.
\textit{IEEE Robotics and Automation Letters} 7(2):5663--5670

\bibitem{alatur2023granular}
Alatur N, Andersson O, Siegwart R, Ott L. 2023.
In \textit{2023 {{IEEE}}/{{RSJ International Conference}} on {{Intelligent
  Robots}} and {{Systems}} ({{IROS}})}

\bibitem{Li2019}
Li Y, Wu J, Tedrake R, Tenenbaum JB, Torralba A. 2018.
\textit{arXiv:1810.01566}

\bibitem{aoyama2024granular}
Aoyama Y, Haeri A, Theodorou EA. 2024.
In \textit{2024 {{IEEE International Conference}} on {{Robotics}} and
  {{Automation}} ({{ICRA}})}

\bibitem{prados2021granular}
Prados A. 2021.
\textit{Physical Review Research} 3(2):023128

\bibitem{ruiz-pino2022granular}
{Ruiz-Pino} N, Prados A. 2022.
\textit{Entropy} 24(1):131

\bibitem{Lin2014}
Lin Q, Tordesillas A. 2014.
\textit{Journal of Industrial and Management Optimization} 10(1):337--362

\bibitem{martinez2016engineered}
Mart{\'\i}nez IA, Petrosyan A, Gu{\'e}ry-Odelin D, Trizac E, Ciliberto S. 2016.
\textit{Nature Physics} 12:843--846

\bibitem{gueryodelin2023driving}
Gu{\'e}ry-Odelin D, Jarzynski C, Plata CA, Prados A, Trizac E. 2023.
\textit{Reports on Progress in Physics} 86:035902

\bibitem{Swan2014}
Swan JW, Bauer JL, Liu Y, Furst EM. 2014.
\textit{Soft Matter} 10(8):1102--1109

\bibitem{Sherman2019}
Sherman ZM, Swan JW. 2019.
\textit{ACS Nano} 13(1):764--771

\bibitem{Jha2011}
Jha PK, Kuzovkov V, Grzybowski BA, de~la Cruz MO. 2011.
\textit{Soft Matter} 8(1):227--234

\bibitem{Long2018}
Long C, Lei Ql, Ren Cl, Ma Yq. 2018.
\textit{The Journal of Physical Chemistry B} 122(12):3196--3201

\bibitem{Klotsa2013}
Klotsa D, Jack RL. 2013.
\textit{The Journal of Chemical Physics} 138(9):094502

\bibitem{Ramaswamy2015}
Ramaswamy S, Barton PI, Stephanopoulos G. 2015.
\textit{Industrial \& Engineering Chemistry Research} 54(34):8520--8532

\bibitem{Banerjee2010}
Banerjee A, Pomerance A, Losert W, Gupta S. 2010.
\textit{IEEE Transactions on Automation Science and Engineering} 7(2):218--227

\bibitem{Xue2013}
Xue Y, Beltran-Villegas DJ, Bevan MA, Grover MA. 2013.
In \textit{2013 {{American Control Conference}}}

\bibitem{Zhang2020a}
Zhang J, Yang J, Zhang Y, Bevan MA. 2020.
\textit{Science Advances} 6(48):eabd6716

\bibitem{Whitelam2020}
Whitelam S, Tamblyn I. 2020.
\textit{Physical Review E} 101(5):052604

\bibitem{Goodrich2021}
Goodrich CP, King EM, Schoenholz SS, Cubuk ED, Brenner MP. 2021.
\textit{Proceedings of the National Academy of Sciences of the United States of
  America} 118(10):e2024083118

\bibitem{King2024}
King EM, Du CX, Zhu QZ, Schoenholz SS, Brenner MP. 2024.
\textit{Proceedings of the National Academy of Sciences of the United States of
  America} 121(27):e2311891121

\bibitem{Falk-2021-Phys.Rev.Res.}
Falk MJ, Alizadehyazdi V, Jaeger H, Murugan A. 2021.
\textit{Physical Review Research} 3(3):033291

\bibitem{Palacci-2013-Science}
Palacci J, Sacanna S, Steinberg AP, Pine DJ, Chaikin PM. 2013.
\textit{Science} 339(6122)

\bibitem{Fily-2012-PhysRevLett}
Fily Y, Marchetti M. 2012.
\textit{Physical Review Letters} 108(23):235702

\bibitem{Yang-2018-ACSNano}
Yang Y, Bevan MA. 2018.
\textit{ACS Nano} 12(11):10712--10724

\bibitem{Frangipane-2018-eLife}
Frangipane G, Dell'Arciprete D, Petracchini S, Maggi C, Saglimbeni F, et~al.
  2018.
\textit{eLife} 7:e36608

\bibitem{Arlt-2018-NatCommun}
Arlt J, Martinez VA, Dawson A, Pilizota T, Poon WCK. 2018.
\textit{Nature Communications} 9(1):768

\bibitem{Massana-Cid-2024-NatCommun}
Massana-Cid H, Maggi C, Gnan N, Frangipane G, Di~Leonardo R. 2024.
\textit{Nature Communications} 15(1):6574

\bibitem{Dervaux-2017-NaturePhys}
Dervaux J, Capellazzi~Resta M, Brunet P. 2017.
\textit{Nature Physics} 13(3):306--312

\bibitem{Massana-Cid-2022-NatCommun}
{Massana-Cid} H, Maggi C, Frangipane G, Di~Leonardo R. 2022.
\textit{Nature Communications} 13(1):2740

\bibitem{Colabrese-2017-Phys.Rev.Lett.}
Colabrese S, Gustavsson K, Celani A, Biferale L. 2017.
\textit{Physical Review Letters} 118(15):158004

\bibitem{Muinos-Landin-2021-Sci.Robot.}
{Mui{\~n}os-Landin} S, Fischer A, Holubec V, Cichos F. 2021.
\textit{Science Robotics} 6(52):eabd9285

\bibitem{Juarez2012}
Juárez JJ, Bevan MA. 2012.
\textit{Advanced Functional Materials} 22(18):3833--3839

\bibitem{Tang2016}
Tang X, Rupp B, Yang Y, Edwards TD, Grover MA, Bevan MA. 2016.
\textit{ACS Nano} 10(7):6791--6798

\bibitem{Gao2019a}
Gao Y, Lakerveld R. 2019.
\textit{AIChE Journal} 65(6):e16582

\bibitem{Baldovin-2023-Phys.Rev.Lett.}
Baldovin M, Guéry-Odelin D, Trizac E. 2023.
\textit{Physical Review Letters} 131(11):118302Publisher: American Physical
  Society

\bibitem{Sanchez-2012-Naturea}
Sanchez T, Chen DTN, DeCamp SJ, Heymann M, Dogic Z. 2012.
\textit{Nature} 491(7424):431--434

\bibitem{Giomi-2014-Philos.Trans.R.Soc.Math.Phys.Eng.Sci.}
Giomi L, Bowick MJ, Mishra P, Sknepnek R, Cristina~Marchetti M. 2014.
\textit{Philosophical Transactions of the Royal Society A: Mathematical,
  Physical and Engineering Sciences} 372(2029):20130365

\bibitem{Ross-2019-Nature}
Ross TD, Lee HJ, Qu Z, Banks RA, Phillips R, Thomson M. 2019.
\textit{Nature} 572(7768):224--229

\bibitem{Najma-2022-NatCommun}
Najma B, Varghese M, Tsidilkovski L, Lemma L, Baskaran A, Duclos G. 2022.
\textit{Nature Communications} 13(1):6465

\bibitem{Lemma-2023-PNASNexus}
Lemma LM, Varghese M, Ross TD, Thomson M, Baskaran A, Dogic Z. 2023.
\textit{PNAS Nexus} 2(5):pgad130

\bibitem{Zhang-2021-NatMater}
Zhang R, Redford SA, Ruijgrok PV, Kumar N, Mozaffari A, et~al. 2021.
\textit{Nature Materials} :1--8

\bibitem{Shankar-2019-Phys.Rev.X}
Shankar S, Marchetti MC. 2019.
\textit{Physical Review X} 9(4):041047

\bibitem{Norton-2020-Phys.Rev.Lett.}
Norton MM, Grover P, Hagan MF, Fraden S. 2020.
\textit{Physical Review Letters} 125(17):178005

\bibitem{Ghosh-2024-a}
Ghosh S, Baskaran A, Hagan MF. 2024.
\textit{arXiv:2408.14596}

\bibitem{ghosh_spatiotemporal_2024}
Ghosh S, Joshi C, Baskaran A, Hagan MF. 2024.
\textit{Soft Matter} 20(35):7059--7071

\bibitem{Sinigaglia-2024-Phys.Rev.Lett.}
Sinigaglia C, Braghin F, Serra M. 2024.
\textit{Physical Review Letters} 132(21):218302

\bibitem{Shankar-2024-Proc.Natl.Acad.Sci.}
Shankar S, Scharrer LVD, Bowick MJ, Marchetti MC. 2024.
\textit{Proceedings of the National Academy of Sciences of the United States of
  America} 121(21):e2400933121

\bibitem{Schuppler-2016-NatCommun}
Schuppler M, Keber FC, Kr{\"o}ger M, Bausch AR. 2016.
\textit{Nature Communications} 7(1):13120

\bibitem{Linsmeier-2016-NatCommuna}
Linsmeier I, Banerjee S, Oakes PW, Jung W, Kim T, Murrell MP. 2016.
\textit{Nature Communications} 7(1):12615

\bibitem{Clarke-2025-a}
Clarke J, Cavanna F, Marne A, Davolio A, Alvarado J. 2025.
\textit{arXiv:2502.18672}

\bibitem{Liu-2025-Nat.Phys.}
Liu J, Burkart T, Ziepke A, Reinhard J, Chao YC, et~al. 2025.
\textit{Nature Physics} :1--10

\bibitem{Blanc-2024-Langmuir}
Blanc B, Zhang Z, Liu E, Zhou N, Dellatolas I, et~al. 2024{\natexlab{a}}.
\textit{Langmuir} 40(13):6862--6868

\bibitem{Blanc-2024-Proc.Natl.Acad.Sci.}
Blanc B, Agyapong JN, Hunter I, Galas JC, {Fernandez-Nieves} A, Fraden S.
  2024{\natexlab{b}}.
\textit{Proceedings of the National Academy of Sciences of the United States of
  America} 121(6):e2313258121

\bibitem{Nelson-2002-NanoLett.}
Nelson DR. 2002.
\textit{Nano Letters} 2(10):1125--1129

\bibitem{Yang-2025-Nat.Mater.}
Yang F, Liu S, Lee HJ, Phillips R, Thomson M. 2025.
\textit{Nature Materials} :1--11

\bibitem{Maroudas-Sacks-2021-Nat.Phys.}
Maroudas-Sacks Y, Garion L, Shani-Zerbib L, Livshits A, Braun E, Keren K. 2021.
\textit{Nature Physics} 17(2):251--259

\bibitem{Saw-2017-Naturea}
Saw TB, Doostmohammadi A, Nier V, Kocgozlu L, Thampi S, et~al. 2017.
\textit{Nature} 544(7649):212--216

\bibitem{Copenhagen-2021-Nat.Phys.}
Copenhagen K, Alert R, Wingreen NS, Shaevitz JW. 2021.
\textit{Nature Physics} 17(2):211--215

\bibitem{Norton-2024-Phys.Rev.E}
Norton MM, Grover P. 2024.
\textit{Physical Review E} 110(5):054605

\bibitem{salamon_thermodynamic_1983}
Salamon P, Berry RS. 1983.
\textit{Physical Review Letters} 51:1127--1130

\bibitem{crooks_measuring_2007}
Crooks GE. 2007.
\textit{Physical Review Letters} 99:100602

\bibitem{sivak_thermodynamic_2012}
Sivak DA, Crooks GE. 2012.
\textit{Physical Review Letters} 108(19):190602

\bibitem{blaber_optimal_2023}
Blaber S, Sivak DA. 2023.
\textit{Journal of Physics Communications} 7(3):033001

\bibitem{tafoya_using_2019}
Tafoya S, Large SJ, Liu S, Bustamante C, Sivak DA. 2019.
\textit{Proceedings of the National Academy of Sciences of the United States of
  America} 116(13):5920--5924

\bibitem{gupta_optimal_2022}
Gupta D, Large SJ, Toyabe S, Sivak DA. 2022.
\textit{Journal of Physical Chemistry Letters} 13(51):11844--11849

\bibitem{loos_universal_2024}
Loos SAM, Monter S, Ginot F, Bechinger C. 2024.
\textit{Physical Review X} 14(2):021032

\bibitem{Davis-2024-Phys.Rev.X}
Davis LK, Proesmans K, Fodor {\'E}. 2024.
\textit{Physical Review X} 14(1):011012

\bibitem{garcia-millan2024optimal}
Garcia-Millan R, Sch{\"u}ttler J, M.~E.~Cates , Loos SAM. 2024.
\textit{arXiv:2407.18542}

\bibitem{schuttler2025active}
Sch{\"u}ttler J, Garcia-Millan R, Cates ME, Loos SAM. 2025.
\textit{arXiv:2501.18613}

\bibitem{Markovich-2021-Phys.Rev.Xa}
Markovich T, Fodor {\'E}, Tjhung E, Cates ME. 2021.
\textit{Physical Review X} 11(2):021057

\bibitem{Fodor-2022-Annu.Rev.Condens.MatterPhys.b}
Fodor {\'E}, Jack RL, Cates ME. 2022.
\textit{Annual Review of Condensed Matter Physics} 13:215--238

\bibitem{Gupta-2023-Phys.Rev.E}
Gupta D, Klapp SHL, Sivak DA. 2023.
\textit{Physical Review E} 108(2):024117

\bibitem{engel_optimal_2022}
Engel MC, Smith JA, Brenner MP. 2022.
\textit{arXiv:2201.00098}

\bibitem{whitelam_demon_2023}
Whitelam S. 2023.
\textit{Physical Review X} 13(2):021005

\bibitem{aurell_optimal_2011}
Aurell E, Mej{\'i}a-Monasterio C, Muratore-Ginanneschi P. 2011.
\textit{Physical Review Letters} 106(25):250601

\bibitem{nakazato_geometrical_2021}
Nakazato M, Ito S. 2021.
\textit{Physical Review Research} 3(4):043093

\bibitem{proesmans_finite-time_2020}
Proesmans K, Ehrich J, Bechhoefer J. 2020.
\textit{Physical Review Letters} 125(10):100602

\bibitem{zhong_beyond_2024}
Zhong A, DeWeese MR. 2024.
\textit{Physical Review Letters} 133(5):057102

\bibitem{Boyden-2005-NatNeurosci}
Boyden ES, Zhang F, Bamberg E, Nagel G, Deisseroth K. 2005.
\textit{Nature Neuroscience} 8(9):1263--1268

\bibitem{ElowitzInternalExternalNoise}
Elowitz MB, Levine AJ, Siggia ED, Swain PS. 2002.
\textit{Science} 297(5584):1183--1186

\bibitem{lestas_fundamental_2010}
Lestas I, Vinnicombe G, Paulsson J. 2010.
\textit{Nature} 467(7312):174--178

\bibitem{aoki_universal_2019}
Aoki SK, Lillacci G, Gupta A, Baumschlager A, Schweingruber D, Khammash M.
  2019.
\textit{Nature} 570(7762):533--537

\bibitem{Yi-2007-MethodsinEnzymology}
Yi T, Andrews BW, Iglesias PA. 2007.
In \textit{Methods in {Enzymology}}, eds. MI~Simon, BR~Crane, A~Crane, vol. 422
  of \textit{Two‐{Component} {Signaling} {Systems}, {Part} {A}}. Academic
  Press,  123--140

\bibitem{Cowan-2014-IntegrativeandComparativeBiology}
Cowan NJ, Ankarali MM, Dyhr JP, Madhav MS, Roth E, et~al. 2014.
\textit{Integrative and Comparative Biology} 54(2):223--237

\bibitem{liu2016control}
Liu YY, Barab\'asi AL. 2016.
\textit{Reviews of Modern Physics} 88:035006

\bibitem{DSouza2023network}
D'Souza RM, Di~Bernardo M, Liu YY. 2023.
\textit{Nature Reviews Physics} 5(4):250--262

\bibitem{Wu2024}
Wu Z, Huang L, Wang M, He X. 2024.
\textit{Psychoradiology} 4:kkae028

\bibitem{Kulkarni2025}
Kulkarni S, Bassett DS. 2025.
\textit{Annual Review of Biophysics} 54

\bibitem{Gu2015}
Gu S, Pasqualetti F, Cieslak M, Telesford QK, Yu AB, et~al. 2015.
\textit{Nature Communications} 6:1--10

\bibitem{Tang2017}
Tang E, Giusti C, Baum GL, Gu S, Pollock E, et~al. 2017.
\textit{Nature Communications} 8(1):1252

\bibitem{Betzel2016a}
Betzel RF, Gu S, Medaglia JD, Pasqualetti F, Bassett DS. 2016.
\textit{Scientific Reports} 6(1):30770

\bibitem{Lee2019a}
Lee B, Kang U, Chang H, Cho KH. 2019.
\textit{iScience} 13:154--162

\bibitem{Luppi2024}
Luppi AI, Singleton SP, Hansen JY, Jamison KW, Bzdok D, et~al. 2024.
\textit{Nature Biomedical Engineering} 8(9):1142--1161

\bibitem{Kim2018a}
Kim JZ, Soffer JM, Kahn AE, Vettel JM, Pasqualetti F, Bassett DS. 2018.
\textit{Nature Physics} 14(1):91--98

\bibitem{Cui2020}
Cui Z, Stiso J, Baum GL, Kim JZ, Roalf DR, et~al. 2020.
\textit{eLife} 9:e53060

\bibitem{Parkes2022}
Parkes L, Kim JZ, Stiso J, Calkins ME, Cieslak M, et~al. 2022.
\textit{Science Advances} 8(50):eadd2185

\bibitem{Gu2017}
Gu S, Betzel RF, Mattar MG, Cieslak M, Delio PR, et~al. 2017.
\textit{NeuroImage} 148:305--317

\bibitem{Tang2022a}
Tang B, Zhang W, Deng S, Liu J, Hu N, et~al. 2022.
\textit{BMC Psychiatry} 22(1):26

\bibitem{Meyer-Base2020}
{Meyer-B{\"a}se} L, Saad F, Tahmassebi A. 2020.
In \textit{Medical {{Imaging}} 2020: {{Biomedical Applications}} in
  {{Molecular}}, {{Structural}}, and {{Functional Imaging}}}, vol. 11317. SPIE

\bibitem{nguyen_organization_2021}
Nguyen M, Qiu Y, Vaikuntanathan S. 2021.
\textit{Annual Review of Condensed Matter Physics} 12(1):273--290

\bibitem{HaganARPC}
Perlmutter JD, Hagan MF. 2015.
\textit{Annual Review of Physical Chemistry} 66(Volume 66, 2015):217--239

\bibitem{trubiano_optimization_2022}
Trubiano A, Hagan MF. 2022.
\textit{Journal of Chemical Physics} 157(24):244901

\bibitem{JohnsonOptimalKineticPathways}
Jhaveri A, Loggia S, Qian Y, Johnson ME. 2024.
\textit{Proceedings of the National Academy of Sciences of the United States of
  America} 121(19):e2403384121

\bibitem{kirk1998optimal}
Kirk DE. 1998.
Optimal control theory: An introduction.
Dover Publications

\bibitem{sekimoto1997complementarity}
Sekimoto K, Sasa SI. 1997.
\textit{J. Phys. Soc. Jpn.} 66:3326--3328

\bibitem{lecunuder2016fast}
{Le Cunuder} A, Mart{\'\i}nez IA, Petrosyan A, Gu{\'e}ry-Odelin D, Trizac E,
  Ciliberto S. 2016.
\textit{Appl. Phys. Lett.} 109:113502

\bibitem{szamel2014self-propelled}
Szamel G. 2014.
\textit{Phys. Rev. E} 90:012111

\bibitem{martin2021statistical}
Martin D, O'Byrne J, Cates ME, Fodor {\'E}, Nardini C, et~al. 2021.
\textit{Phys. Rev. E} 103:032607

\bibitem{saha2023information}
Saha TK, Ehrich J, Gavrilov M, Still S, Sivak DA, Bechhoefer J. 2023.
\textit{Physical Review Letters} 131:057101

\bibitem{gomezMarin2008optimal}
Gomez-Marin A, Schmiedl T, Seifert U. 2008.
\textit{J. Chem. Phys.} 129:024114

\end{thebibliography}
\bibliographystyle{ar-style4}

\end{document}